\documentclass[a4paper,conference]{IEEEtran}

\usepackage{cite, multicol}
\usepackage{graphicx} 
\usepackage{subfigure}
\usepackage{amsmath}  

\usepackage{fancyheadings}
\begin{document}

\title{Hyperbolic Divergence Cleaning for SPH}

\author{\IEEEauthorblockN{Terrence S. Tricco}
\IEEEauthorblockA{Monash Centre for Astrophysics\\
Monash University\\
Melbourne, Australia\\
terrence.tricco@monash.edu}
\and
\IEEEauthorblockN{Daniel J. Price}
\IEEEauthorblockA{Monash Centre for Astrophysics\\
Monash University\\
Melbourne, Australia\\
daniel.price@monash.edu}}


\maketitle

\begin{abstract}
We present SPH formulations of Dedner et al's hyperbolic/parabolic divergence cleaning scheme for magnetic and velocity fields.  Our implementation preserves the conservation properties of SPH which is important for stability.  This is achieved by deriving an energy term for the $\psi$ field, and imposing energy conservation on the cleaning subsystem of equations. This necessitates use of conjugate operators for $\nabla \cdot {\bf B}$ and $\nabla \psi$ in the numerical equations.  For both the magnetic and velocity fields, the average divergence error in the system is reduced by an order of magnitude with our cleaning algorithm.  Divergence errors in SPMHD are maintained to $< 1\%$, even for realistic 3D applications with a corresponding gain in numerical stability.  Density errors for an oscillating elliptic water drop using weakly compressible SPH are reduced by a factor of two.
\end{abstract}

\section{Introduction}

Magnetic fields have the property of being divergence free, that is $\nabla \cdot {\bf B} = 0$.  Incompressible fluids have a similar divergence free property for the velocity field.  Maintaining these divergence constraints is one of the central difficulties in performing accurate simulations of magnetohydrodynamics (MHD) and incompressible fluid behaviour.  For MHD in particular, the presence of magnetic monopoles introduces a spurious force which, when large, is disruptive to the dynamics of the system.

Similar approaches can be utilised to satisfy the divergence constraints in both cases.  For example, projection methods construct a divergence free vector field via the solution of a Poisson equation and have been applied successfully to both systems.  Specialised approaches have also been developed for each case.  One example is the constrained transport method \cite{1988ApJ...332..659E} for MHD, which by conserving magnetic flux through a closed surface, can keep the divergence constraint to within machine precision. For SPH simulations of incompressible fluids, a stiff equation of state can be used to limit density variations to $\sim 1\%$ \cite{1994JCoPh.110..399M}, creating a weakly compressible fluid approximating incompressibility.

The hyperbolic divergence cleaning method of Dedner et al \cite{2002JCoPh.175..645D} was introduced for maintaining the $\nabla \cdot {\bf B} = 0$ constraint in MHD.  It involves the addition of a new scalar field, $\psi$, which is coupled to the magnetic field by
\begin{equation}
 \left(\frac{{\rm d}{\bf B}}{{\rm d}t}\right)_\psi = - \nabla \psi .
\label{eq:cleaning-gradpsi}
\end{equation}
This $\psi$ field evolves according to
\begin{equation}
 \frac{{\rm d}\psi}{{\rm d}t} = - c_h^2 \nabla \cdot {\bf B} - \frac{\psi}{\tau} ,
\label{eq:cleaning-psievolve}
\end{equation}
and combined these produce a damped wave equation
\begin{equation}
 \frac{\partial^{2} (\nabla \cdot {\bf B})}{\partial t^2} - c_{h}^{2} \nabla^2 (\nabla \cdot {\bf B}) + \frac{1}{\tau} \frac{\partial (\nabla \cdot {\bf B})}{\partial t} = 0 .
\label{eq:waveeqn}
\end{equation}
Thus divergence is spread away from sources by a series of damped waves. The wave speed, $c_h$, is typically chosen to be the fastest wave obeying the Courant stability condition.  The damping timescale, $\tau$, acts as a diffusion on the divergence.  By using waves to spread the divergence over a larger volume, the amplitude of any single large source is diminished and the diffusion is more effective.  While originally proposed for use on the magnetic field for MHD simulations, this approach would be valid for any vector field.  The damping timescale is set to $\tau^{-1} \equiv \sigma c_{h} / h$, where $h$ is the smoothing length and $\sigma$ is a dimensionless quantity specifying the damping strength.

Hyperbolic divergence cleaning has found popular use in both Eulerian \cite{2010JCoPh.229.2117M, 2009ApJ...696...96W} and Lagrangian based codes \cite{2011MNRAS.414..129G, 2011MNRAS.tmp.1536P}, chiefly for its simplicity, easy implementation, and low computational cost.  However, for the SPH implementation of MHD (SPMHD), this method has not been widely adopted.  Initial implementation attempts by Price \cite{pm05} found divergence reductions were not substantial (a factor $\sim 2$), and the method risked actually increasing divergence in certain test cases.

The work presented here describes a new formulation of hyperbolic divergence cleaning for SPH that removes previous difficulties \cite{tp12}.  Implementations for both the magnetic and velocity fields are presented.  Our formulation imposes the constraint of energy conservation on the subsystem of cleaning equations, guaranteeing that energy transferred to the $\psi$ field must either be conserved or dissipated.  This prevents increases in divergence.

The paper is laid out as follows:  Sec.~\ref{sec:cleaning} discusses hyperbolic divergence cleaning for the magnetic field of SPMHD.  A brief description of SPMHD is presented (Sec.~\ref{sec:spmhd}), along with the Euler Potentials (Sec.~\ref{sec:ep}) and artificial resistivity (Sec.~\ref{sec:resis}) since they will be used as a basis of comparison for the new divergence cleaning method.  In Sec.~\ref{sec:clean-mhd}, the energy contained in the $\psi$ field is derived and modifications are made to the cleaning equations to conserve energy, then the energy conserving SPMHD implementation is constructed (Sec.~\ref{sec:clean-spmhd}).  Hyperbolic divergence cleaning for the velocity field is discussed in Sec.~\ref{sec:velclean}.  Starting from an outline of weakly compressible SPH (Sec.~\ref{sec:wcsph}), a new energy term is created for the $\psi$ field for contributions from the velocity field (Sec.~\ref{sec:velclean-continuum}) which is used to create the conservative SPH implementation (Sec.~\ref{sec:velclean-wcsph}).  Tests of our method are presented in Sec.~\ref{sec:tests}, applied to three MHD problems and one incompressible fluid problem.  The SPMHD tests include a simple free boundary test (Sec.~\ref{sec:free-boundary}), the Orszag-Tang vortex where the cleaning method is compared against resistivity and Euler Potentials (Sec.~\ref{sec:ot}), and a collapsing molecular cloud involving star formation (Sec.~\ref{sec:star}).  A test of the velocity cleaning is presented on an oscillating elliptic water drop using weakly compressible SPH (Sec.~\ref{sec:drop}).  Conclusions are presented in Sec.~\ref{sec:conclusion}.


\section{Hyperbolic divergence cleaning for the MHD equations}
\label{sec:cleaning}

\subsection{Smoothed particle magnetohydrodynamics}
\label{sec:spmhd}

The equations of ideal MHD solved in SPH are given by
\begin{align}
\rho_a =& \sum_b m_b W_{ab} (h_a), \label{eq:sphcty} \\
h_{a} =& \eta \left( \frac{m_{a}}{\rho_{a}}\right)^{1/n_\text{dim}}, \label{eq:sml} \\
\frac{{\rm d}{\bf{v}}_a}{{\rm d}t} = 
& \sum_b m_b \bigg[\frac{{\bf M}_a}{\Omega_a \rho_a^2}\cdot \nabla_a W_{ab}(h_a) \nonumber \\
& \hspace{19mm} + \frac{{\bf M}_{b}}{\Omega_b \rho_b^2} \cdot \nabla_b W_{ab}(h_b) \bigg], \label{eq:spmhd-momentum-eqn} \\
 \frac{{\rm d}{\bf{B}}_a}{{\rm d}t} =& - \frac{1}{\Omega_a \rho_a} \sum_b m_b \bigg[ {\bf{v}}_{ab} \left( {\bf{B}}_a \cdot \nabla_a W_{ab}(h_a) \right) \nonumber \\
& \hspace{19mm}- {\bf{B}}_a \left( {\bf{v}}_{ab} \cdot \nabla_a W_{ab}(h_a) \right) \bigg], \label{eq:sphind}
\end{align}
Here, ${\bf v}$ is the fluid velocity, ${\bf B}$ is the magnetic field, and ${\rm d}/{\rm d}t$ is the material derivative.  The density, $\rho$, is calculated via summation using an iterative procedure to self consistently determine the smoothing length, $h$. Variable smoothing length gradients are accounted for with the $\Omega$ terms (see \cite{sh02}).  The momentum equation contains contributions from thermal pressure, $P$, and the Lorentz force, given in terms of the Maxwell stress tensor
\begin{equation}
{\bf M} = {\bf B}{\bf B} - \left(P + \tfrac{1}{2} B^2\right) {\bf I} .
\label{eq:divM}
\end{equation}
The contribution from any spurious ${\bf B} (\nabla \cdot {\bf B})$ force is subtracted out by including the additional term
\begin{align}
 \left(\frac{{\rm d}{\bf{v}}_a}{{\rm d}t}\right)_{\nabla \cdot \bf{B}} = - {\bf B}_a \sum_b m_b \bigg[ &\frac{{\bf B}_a}{\Omega_a \rho_a^2} \cdot \nabla_a W_{ab}(h_a) \nonumber \\
+ &\frac{{\bf B}_b}{\Omega_b \rho_b^2} \cdot \nabla_a W_{ab}(h_b) \bigg] .
\label{eq:tensile-instability-correction}
\end{align} 
The induction equation is derived from ${\partial {\bf B}} / {\partial t} = \nabla \times ({\bf v} \times {\bf B})$ with the monopole contribution removed.  Hence, this scheme is formally equivalent to Powell's eight wave approach \cite{1994arsm.rept.....P}.

\subsubsection{Euler Potentials}
\label{sec:ep}

One method for maintaining the divergence constraint on the magnetic field is to use the Euler Potentials, defining ${\bf B} = \nabla \alpha \times \nabla \beta$.  The potentials are advected exactly, representing the field lines being frozen to the fluid, removing the need to solve the induction equation (\ref{eq:sphind}).  This has had reasonable success controlling divergence error to $\sim 1\%$ (ie, \cite{2009MNRAS.397..733K, 2007MNRAS.377...77P}), however the Euler Potentials place limitations on the possible field configurations which can be represented.  Magnetic field windings in particular cannot be modelled past one rotation, and such topologies would be anticipated for many astrophysical problems of interest.

\subsubsection{Artificial Resistivity}
\label{sec:resis}

Artificial resistivity is added to SPMHD to capture magnetic shocks and discontinuities.  It is similar to artificial viscosity, with form
\begin{equation}
 \frac{{\rm d}{\bf B}}{{\rm d}t} = \rho_a \sum_b m_b \frac{\alpha_B v_\text{sig}}{\overline{\rho}_{ab}} ({\bf B}_a - {\bf B}_b) \hat{{\bf r}} \cdot \nabla_a W_{ab} ,
\end{equation}
where $\alpha_B$ is a dimensionless quantity of order unity, $v_\text{sig}$ is a signal velocity, and $\overline{\rho}_{ab}$ is the average density between particles $a$ and $b$.  This is representative of real resistivity,
\begin{equation}
 \frac{{\rm d}{\bf B}}{{\rm d}t} = \eta \nabla \left(\nabla \cdot \bf{B}\right)  - \eta \nabla  \times ( \nabla \times {\bf B}) ,
\end{equation}
and as such provides diffusion of magnetic divergence.  In some cases, this may be sufficient to control errors, however it is a poor tool to control divergence error since it also dissipates the physical portions of the field.

\subsection{Hyperbolic magnetic divergence cleaning}
\label{sec:clean-mhd}

If just the cleaning system of equations given by (\ref{eq:cleaning-gradpsi}) and (\ref{eq:cleaning-psievolve}) are considered, then the total energy can be written as 
\begin{equation}
 E = \int \left[ \frac{B^2}{2 \mu_0 \rho} + e_\psi \right] \rho {\rm d}V ,
\end{equation}
which is the sum of magnetic energy and as yet undetermined energy contained in the $\psi$ field.  By conservation of energy,
\begin{equation}
\frac{{\rm d}E}{{\rm d}t} = \int \left[ \frac{\bf B}{\mu_0 \rho} \cdot \left( \frac{{\rm d} \bf B}{{\rm d}{t}} \right)_\psi + \frac{{\rm d} e_\psi}{{\rm d}t} \right]  \rho {\rm d}V = 0,
\label{eq:mhd-energyconserv}
\end{equation}
and if $e_\psi$ is assumed to have differentiable form
\begin{equation}
 \frac{{\rm d} e_\psi}{{\rm d}t} = \chi \frac{{\rm d}\psi}{{\rm d}t} ,
\end{equation}
then by inserting (\ref{eq:cleaning-gradpsi}) and (\ref{eq:cleaning-psievolve}), we can obtain
\begin{equation}
 \int \left[ - \frac{{\bf B}}{\mu_0 \rho} \cdot \nabla \psi - \chi c_h^2 \nabla \cdot {\bf B} \right] \rho {\rm d}V = 0 .
\label{eq:asdf}
\end{equation}

Integrating the first term of (\ref{eq:asdf}) by parts will yield
\begin{equation}
 \int \left[ \frac{\psi}{\mu_0\rho}  - \chi c_h^2 \right] (\nabla \cdot {\bf{B}}) \rho {\rm d}V -  \frac{1}{\mu_0} \int_s \psi {\bf{B}} \cdot {\rm d}\hat{\bf{s}} = 0.
\end{equation}
The surface integral may be ignored.  Similar terms appear, for example, in the SPH continuity equation which are likewise taken to be zero \cite{price08}. The remaining term thus implies \mbox{$\chi = \psi / \mu_0 \rho c_h^2$} and therefore
\begin{equation}
 e_\psi = \frac{\psi^2}{2 \mu_0 \rho c_h^2} .
\label{eq:epsi}
\end{equation}

Inserting this energy term into (\ref{eq:mhd-energyconserv}) will produce
\begin{equation}
\label{eq:psi_energy_density_variation}
\int \left[ \frac{\bf B}{\mu_0 \rho} \cdot \left( \frac{{\rm d} \bf B}{{\rm d}{t}} \right)_\psi + \frac{\psi}{\mu_0 \rho c_h^2} \frac{{\rm d}\psi}{{\rm d}t} - \frac{\psi^2}{2 \mu_0 \rho_a^2 c_h^2} \frac{{\rm d}\rho}{{\rm d}t} \right]  \rho {\rm d}V = 0.
\end{equation}
From the preceding analysis, it is clear that energy changes from the first two terms will be balanced by each other.  To account for the third term, the evolution equation for $\psi$ can modified to
\begin{equation}
 \frac{{\rm d} \psi}{{\rm d} t} = -c_h^2 \nabla \cdot {\bf{B}} - \frac{\psi}{\tau} - \tfrac{1}{2} \psi \nabla \cdot {\bf v} ,
\label{eq:psi_evolution_halfdivv} 
\end{equation}
replacing (\ref{eq:cleaning-psievolve}).

\subsection{Discretised hyperbolic magnetic divergence cleaning}
\label{sec:clean-spmhd}

Hyperbolic divergence cleaning is implemented into SPMHD using the differenced derivative operator for $\nabla \cdot {\bf B}$,
\begin{equation}
 \nabla \cdot {\bf B}_a = - \frac{1}{\Omega_a \rho_a} \sum_b m_b \left({\bf B}_a - {\bf B}_b \right) \cdot \nabla W_{ab}(h_a) .
\label{eq:divb}
\end{equation}
Other operator choices are permissible.  It may seem that using the same operator as in the momentum equation (see (\ref{eq:tensile-instability-correction})) would be desirable, however we have found that doing so leads to excessive magnetic energy dissipation.  This occurs because that operator also measures the disorder in the particle arrangement, which the cleaning method attempts to compensate for by adjusting the magnetic field.

The SPMHD analogue of (\ref{eq:mhd-energyconserv}) is
\begin{equation}
\sum_a m_a \left[ \frac{{\bf B}_a}{\mu_0 \rho_{a}} \cdot \left( \frac{{\rm d}{\bf B}_a}{{\rm d}t} \right)_\psi + \frac{\psi_a}{\mu_0 \rho_{a} c_h^2} \frac{{\rm d}\psi_a}{{\rm d}t} \right ] = 0 .
\label{eq:dedtspmhd}
\end{equation}
where the $\psi$ energy term (\ref{eq:epsi}) has been used.  Inserting (\ref{eq:cleaning-psievolve}), with no damping and (\ref{eq:divb}) as the operator choice for $\nabla \cdot {\bf B}$, produces 
\begin{multline}
\sum_a m_a \frac{{\bf B}_a}{\mu_0 \rho_a} \cdot \left(\frac{{\rm d}{\bf B}_a}{{\rm d}t}\right)_\psi = \\
 - \sum_a m_a \frac{\psi_a}{\mu_0 \rho_a^2 \Omega_a} \sum_b m_b \left( {\bf B}_a - {\bf B}_b \right) \cdot \nabla_a W_{ab}(h_a) .
\end{multline}
Splitting the RHS into two halves, performing a change of summation indices on the second half, and recombining, we can obtain
\begin{align}
\left( \frac{{\rm d}{\bf B}_a}{{\rm d}t} \right)_\psi = -\rho_a \sum_b m_b \big[ &\frac{\psi_a}{\Omega_a \rho_a^2} \nabla_a W_{ab}(h_a) \nonumber \\ 
+ &\frac{\psi_b}{\Omega_b \rho_b^2} \nabla_a W_{ab}(h_b) \big].
\label{eq:gradpsisym}
\end{align}

This symmetric form for $\nabla \psi$ is the same as the gradient operator in the momentum equation.  Alternatively, if this had been used as the operator for $\nabla \cdot {\bf B}$, then the differenced derivative operator would be imposed for $\nabla \psi$.  The occurrence of conjugate operators in SPH has been previously noted \cite{cr99}.

The energy change due to damping can be written as
\begin{align}
\left(\frac{{\rm d}E}{{\rm d}t}\right)_{\rm damp} =& \sum_a m_a \frac{\psi_a}{\mu_0 \rho_a c_h^2} \left( \frac{{\rm d}\psi_a}{{\rm d}t} \right)_{\text{damp}} \nonumber \\
=& - \sum_a m_a \frac{\psi_a^2}{\mu_0 \rho_a c_h^2 \tau} ,
\end{align}
which is negative definite.  This guarantees energy may only be removed.

Finally, since the additional $\tfrac{1}{2} \psi (\nabla \cdot {\bf v})$ term introduced to the $\psi$ evolution equation is derived using the continuity equation, the form for $\nabla \cdot {\bf v}$ should be that as in the SPH continuity equation.  Hence,
\begin{equation}
 - \tfrac{1}{2} \psi_a (\nabla \cdot {\bf v}_a) = \frac{\psi_a}{2\Omega_a\rho_a} \sum_b m_b ({\bf v}_a - {\bf v}_b) \cdot \nabla_a W_{ab} (h_{a}).
\end{equation}

\section{Velocity divergence cleaning for weakly compressible SPH}
\label{sec:velclean}

\subsection{Weakly compressible SPH}
\label{sec:wcsph}

A common method for modelling incompressible fluid behaviour with SPH is to use a stiff equation of state with the standard Lagrangian SPH formulation.  This sacrifices true incompressibility for simplicity of implementation.  However, this does not imply computational efficiency as the high speed of sound ($\sim 10\times$ maximum fluid velocity as a minimum) necessitates small sized time steps for stability.  Using the equation of state
\begin{equation}
 P = \frac{c_s^2 \rho_0}{7} \left( \left(\frac{\rho}{\rho_0}\right)^7 - 1\right) ,
\label{eq:wcsph-eos}
\end{equation}
where $\rho_0$ is the reference density of the fluid and $c_s$ is sound speed, this typically results in density variations of $\sim 1\%$ \cite{1994JCoPh.110..399M}.

The equations of motion which are solved are
\begin{equation}
\frac{{\rm d}{\bf v}_a}{{\rm d}t} = - \sum_b m_b \left( \frac{P_a}{\rho_a^2} + \frac{P_b}{\rho_b^2} \right) \nabla_a W_{ab} .
\label{eq:wcsph-momentum}
\end{equation}
In this case, we evolve the density using the SPH equivalent of the continuity equation, 
\begin{equation}
 \frac{{\rm d}\rho_a}{{\rm d}t} = - \sum_b m_b \left({\bf v}_a - {\bf v}_b\right) \nabla_a W_{ab} ,
\label{eq:wcsph-continuity}
\end{equation}
rather than by summation.  The smoothing length of the particles is held constant, calculated according to (\ref{eq:sml}).

\subsection{Hyperbolic divergence cleaning for the velocity field}
\label{sec:velclean-continuum}

Since the continuity equation relies on $\nabla \cdot {\bf v}$ to evolve density, minimising this quantity should lead to improvements in the representation of incompressibility.  We now construct a formulation of divergence cleaning suitable for the velocity field.  The cleaning equations to be solved are modified to become
\begin{align}
\frac{{\rm d}{\bf v}}{{\rm d}t} =& - \frac{\nabla \psi}{\rho} , \label{eq:velclean-gradpsi}\\
\frac{{\rm d}\psi}{{\rm d}t} =& - c_h^2 \rho \nabla \cdot {\bf v} - \frac{\psi}{\tau} . \label{eq:velclean-psievolve}
\end{align}
As the intended application is for incompressible fluids, we assume throughout this section that the density is uniform and constant.  Equations (\ref{eq:velclean-gradpsi}) and (\ref{eq:velclean-psievolve}) still combine to produce the damped wave equation of (\ref{eq:waveeqn}).

We follow a procedure in step with that of Sec.~\ref{sec:clean-mhd}.  The total energy of the velocity-cleaning subsystem is
\begin{equation}
 E = \int \left[ \frac{v^2}{2} + \tilde{e}_\psi \right] \rho {\rm d}V ,
\label{eq:velclean-energyeq}
\end{equation}
and by the constraint of energy conservation
\begin{equation}
 \frac{{\rm d}E}{{\rm d}t} = \int \left[ {\bf v}\frac{{\rm d}{\bf v}}{{\rm d}t} + \chi \frac{{\rm d}\psi}{{\rm d}t} \right] \rho {\rm d}V = 0.
\end{equation}
Inserting (\ref{eq:velclean-gradpsi}) and (\ref{eq:velclean-psievolve}) yields
\begin{equation}
 \int \left[ - {\bf v} \cdot \frac{\nabla \psi}{\rho} - \chi c_h^2 \rho \nabla \cdot {\bf v} \right] \rho {\rm d} V = 0 .
\end{equation}
Integrating the first term by parts, we obtain
\begin{equation}
 \int \left[ \frac{\psi}{\rho} - \chi c_h^2 \rho \right] (\nabla \cdot {\bf v}) \rho {\rm d}V + \int_s \psi {\bf v} \cdot {\rm d}\hat{s} = 0 ,
\end{equation}
which leads to $\chi = \psi / c_h^2 \rho^2$ and hence
\begin{equation}
 \tilde{e}_\psi = \frac{\psi^2}{2 c_h^2 \rho^2} .
\end{equation}

\subsection{Discretised hyperbolic velocity divergence cleaning}
\label{sec:velclean-wcsph}

With the appropriate energy term for this cleaning system, the constrained SPH implementation may be constructed.  We clean using the same $\nabla \cdot {\bf v}$ operator as in the continuity equation, that is,
\begin{equation}
 \nabla \cdot {\bf v}_a = - \frac{1}{\rho_a} \sum_b m_b {\bf v}_{ab} \cdot \nabla_a W_{ab} .
\label{eq:divv}
\end{equation}
The SPH discretised version of (\ref{eq:velclean-energyeq}) is 
\begin{equation}
 E = \sum_a m_a \left[ \frac{v_a^2}{2} + \frac{\psi_a^2}{2 c_h^2 \rho_a^2} \right] .
\end{equation}
Differentiating with respect to time and using (\ref{eq:velclean-psievolve}) and (\ref{eq:divv}), we obtain
\begin{equation}
 \sum_a m_a {\bf v}_a \frac{{\rm d}{\bf v}_a}{{\rm d}t} = \sum_a \frac{m_a \psi_a}{\rho_a^2} \sum_b m_b {\bf v}_{ab} \cdot \nabla_a W_{ab} .
\end{equation}
By splitting the RHS into two halves, swapping summations on one half, then combining, it is concluded that
\begin{equation}
 \frac{{\rm d}{\bf v}_a}{{\rm d}t} = - \sum_b m_b \left( \frac{\psi_a}{\rho_a^2} + \frac{\psi_b}{\rho_b^2} \right) \nabla_a W_{ab} .
\end{equation}
As before, conjugate operators for $\nabla \cdot {\bf v}$ and $\nabla \psi$ become imposed.  In addition to exactly conserving energy, this form for $\nabla \psi$ also conserves momentum.

\section{Tests}
\label{sec:tests}

\subsection{Static cleaning test: free boundaries}
\label{sec:free-boundary}

\begin{figure}
\centering
 \includegraphics[width=\linewidth]{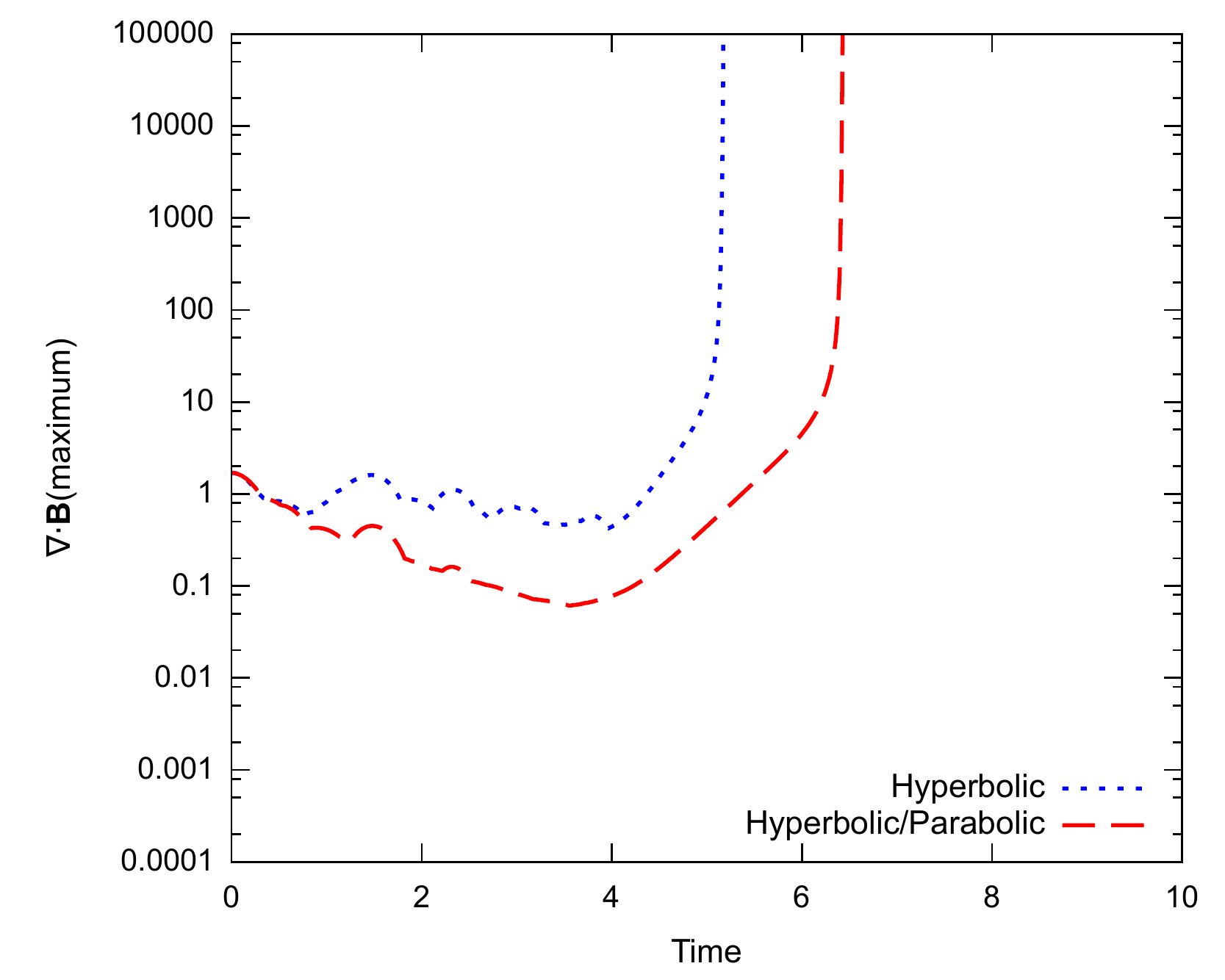} 
 \includegraphics[width=\linewidth]{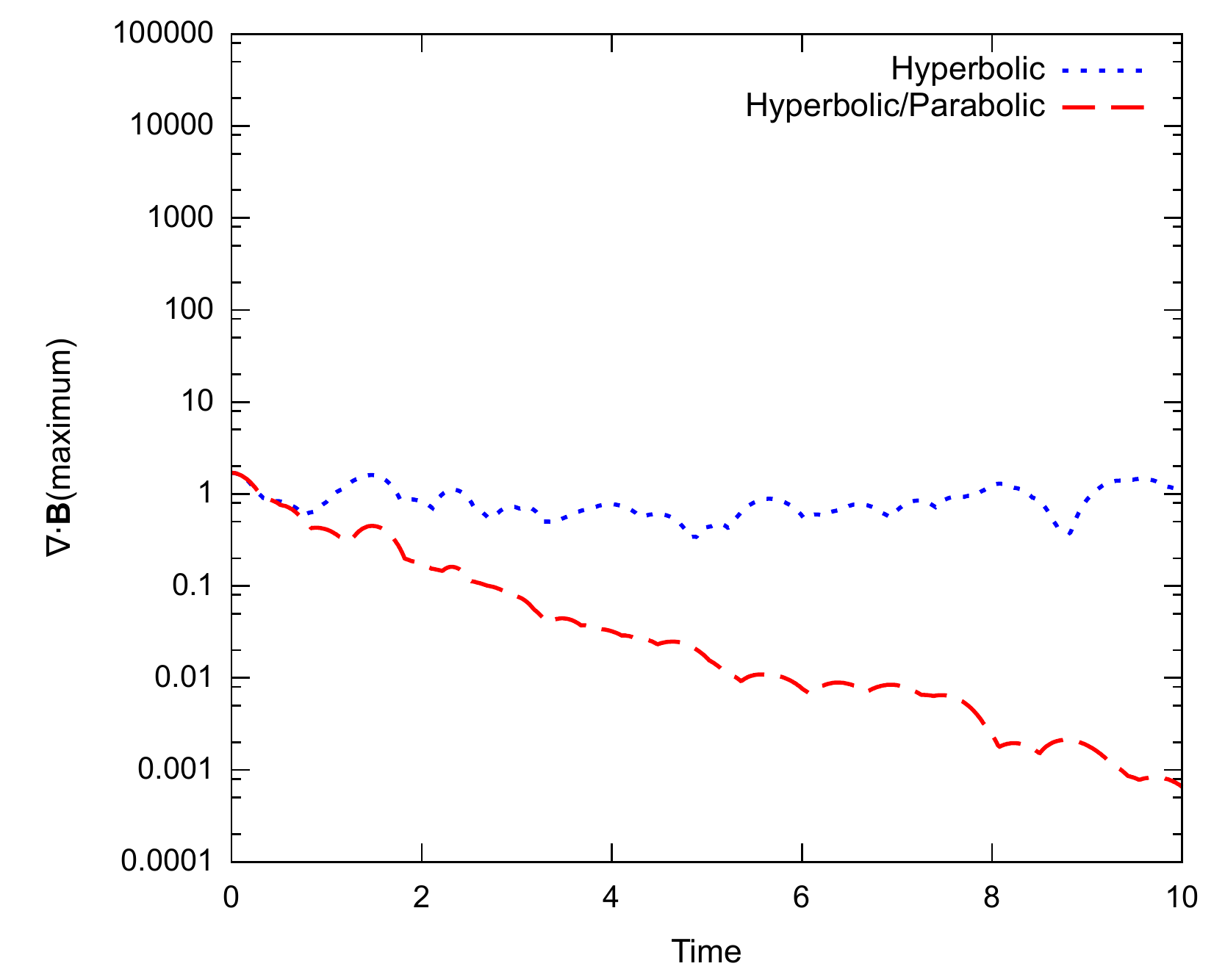}
\caption{Maximum divergence for the free boundary test for the non-conservative formulation (top) and the new constrained divergence cleaning (bottom).  For the non-conservative case, the hard boundary edge acts like an amplifier of divergence causing exponential growth.  With the constrained formulation, the interaction with the boundary is treated correctly and remains stable.}
\label{fig:leftright-div-plots}
\end{figure}

\begin{figure*}
\setlength{\tabcolsep}{0.005\textwidth}
\begin{tabular}{ccccl}
{\scriptsize Control} & {\scriptsize Resistivity} & {\scriptsize Euler Potentials} & {\scriptsize Divergence Cleaning} & \\
   \includegraphics[height=0.22\textwidth]{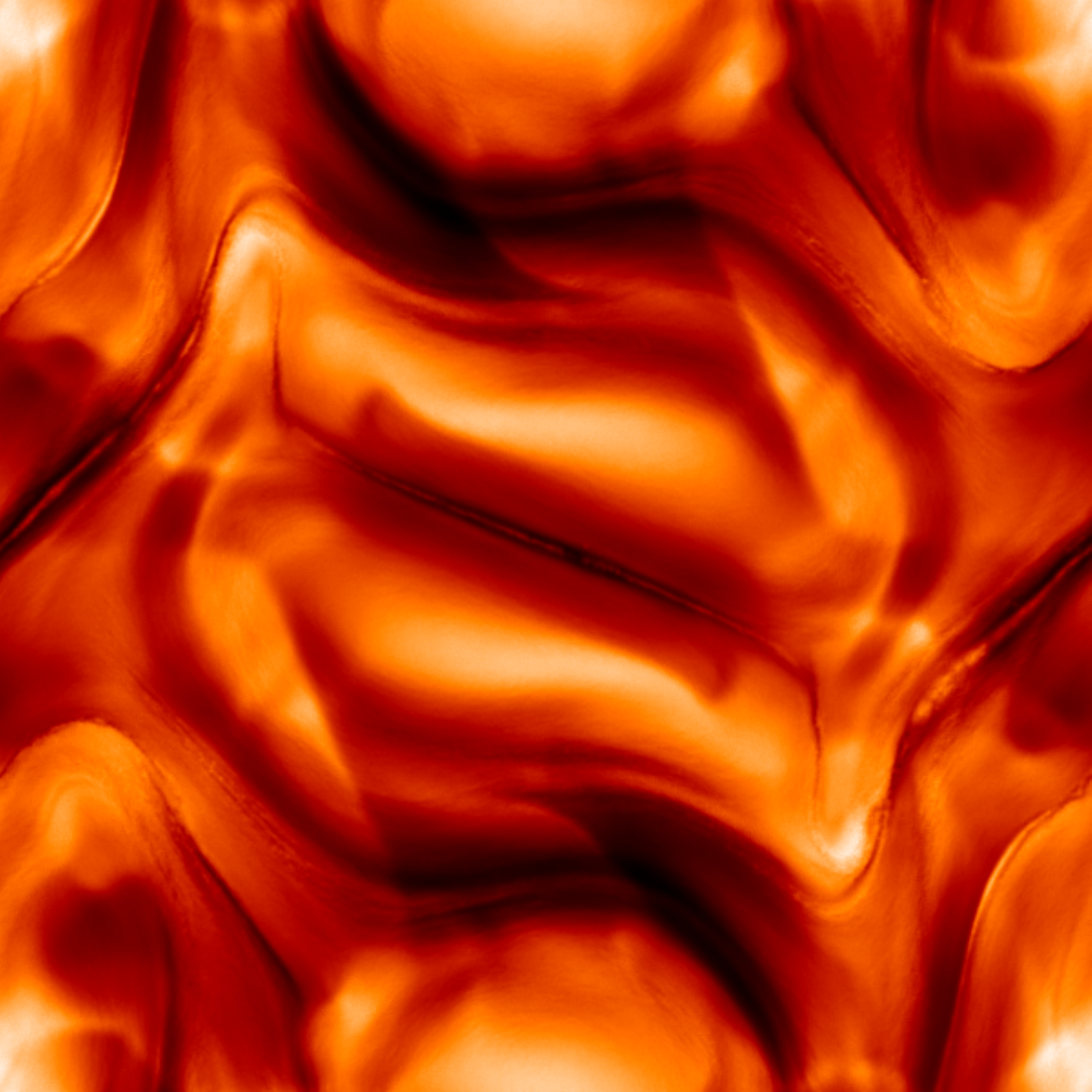} 
 & \includegraphics[height=0.22\textwidth]{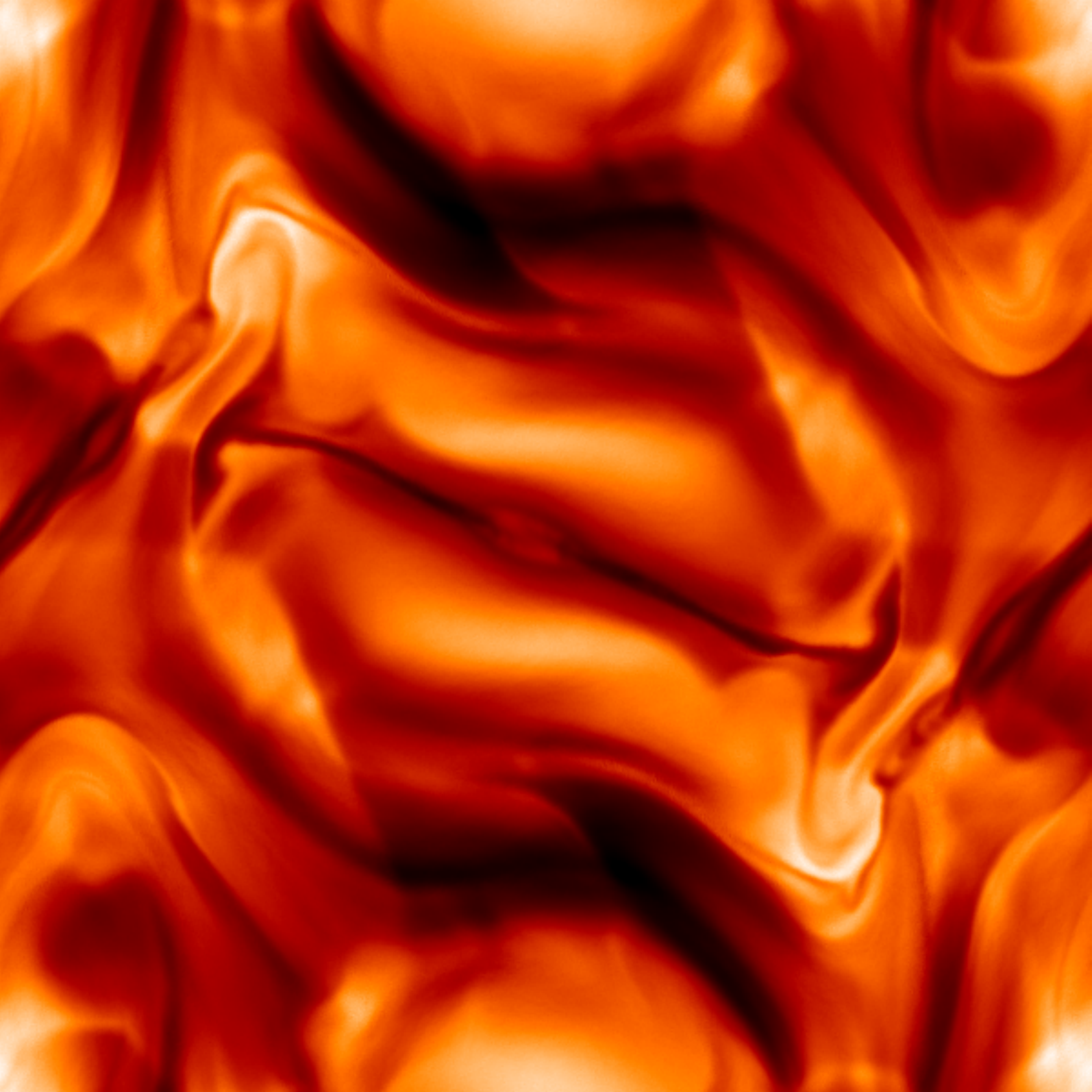}
 & \includegraphics[height=0.22\textwidth]{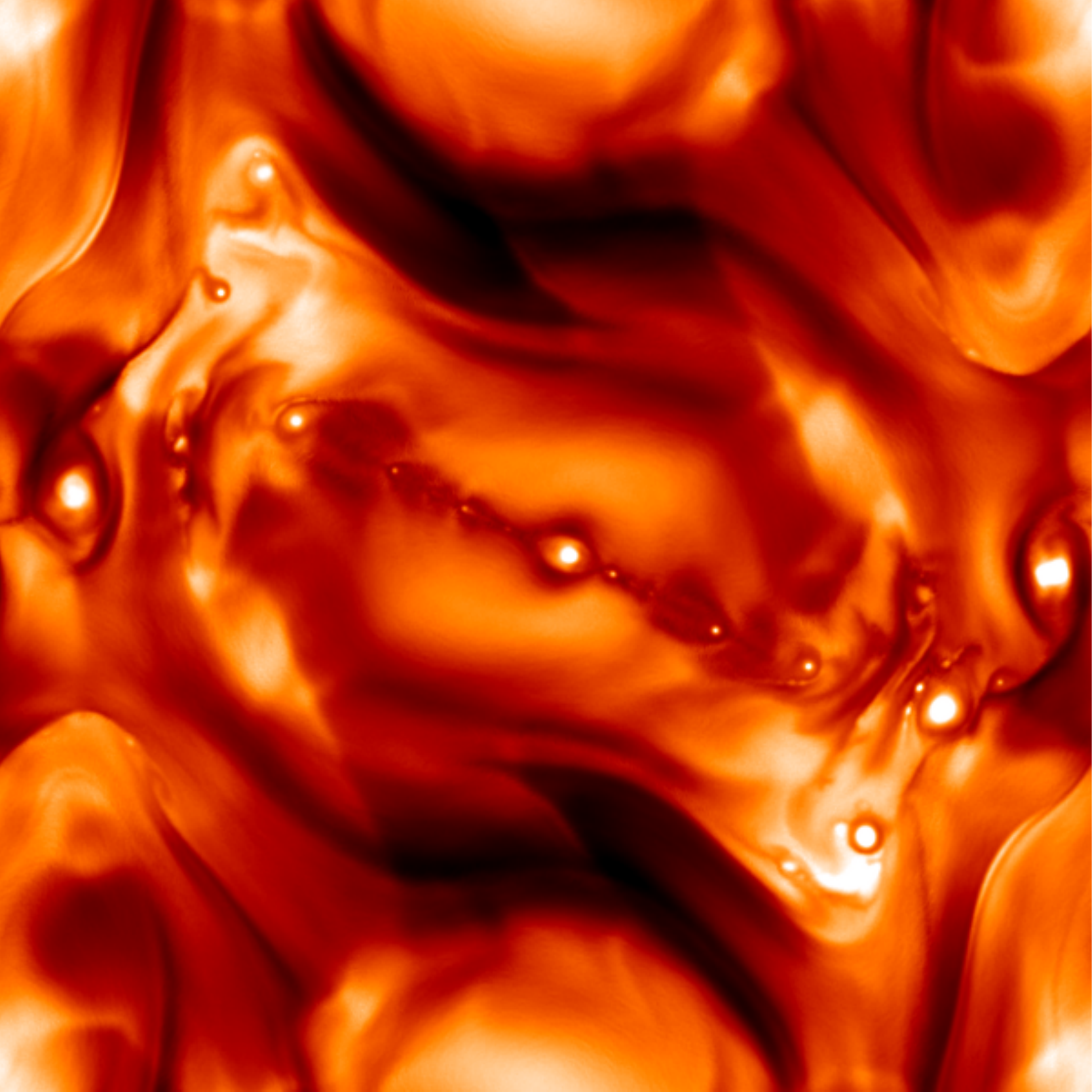}
 & \includegraphics[height=0.22\textwidth]{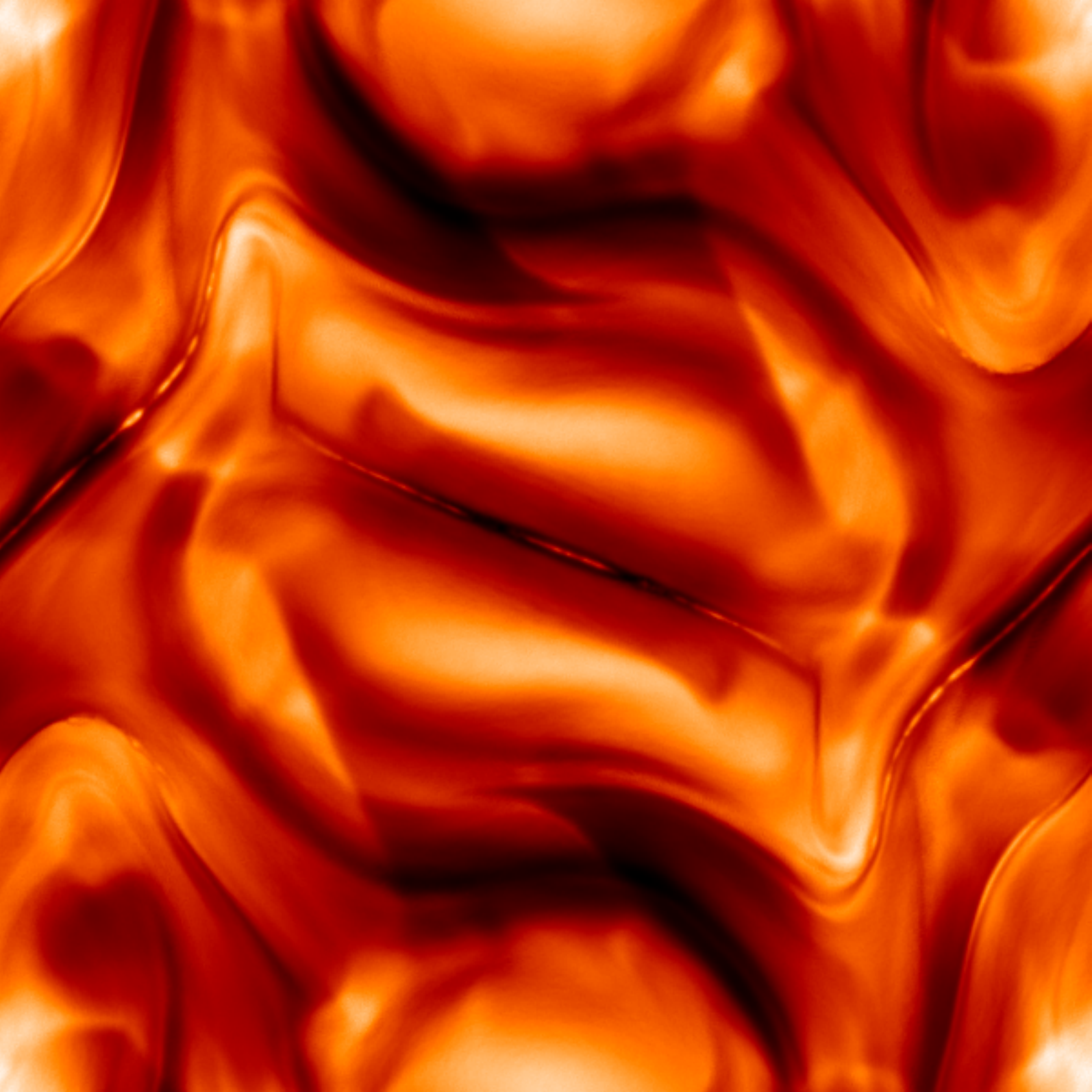}
 & \includegraphics[height=0.22\textwidth]{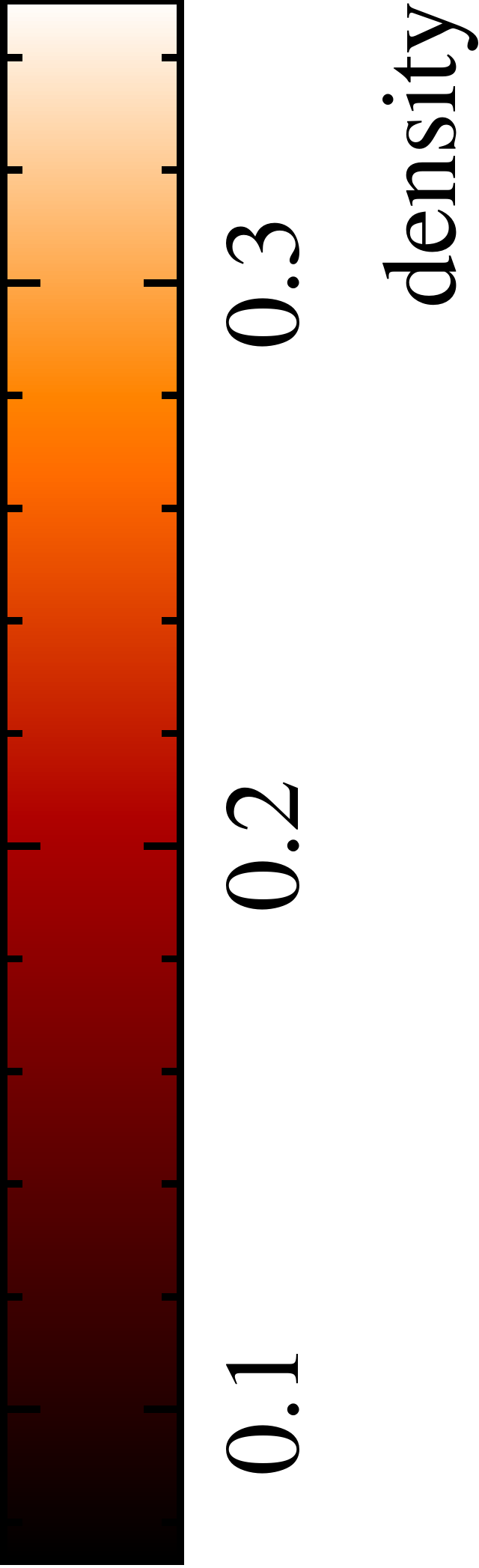}
\\
   \includegraphics[height=0.22\textwidth]{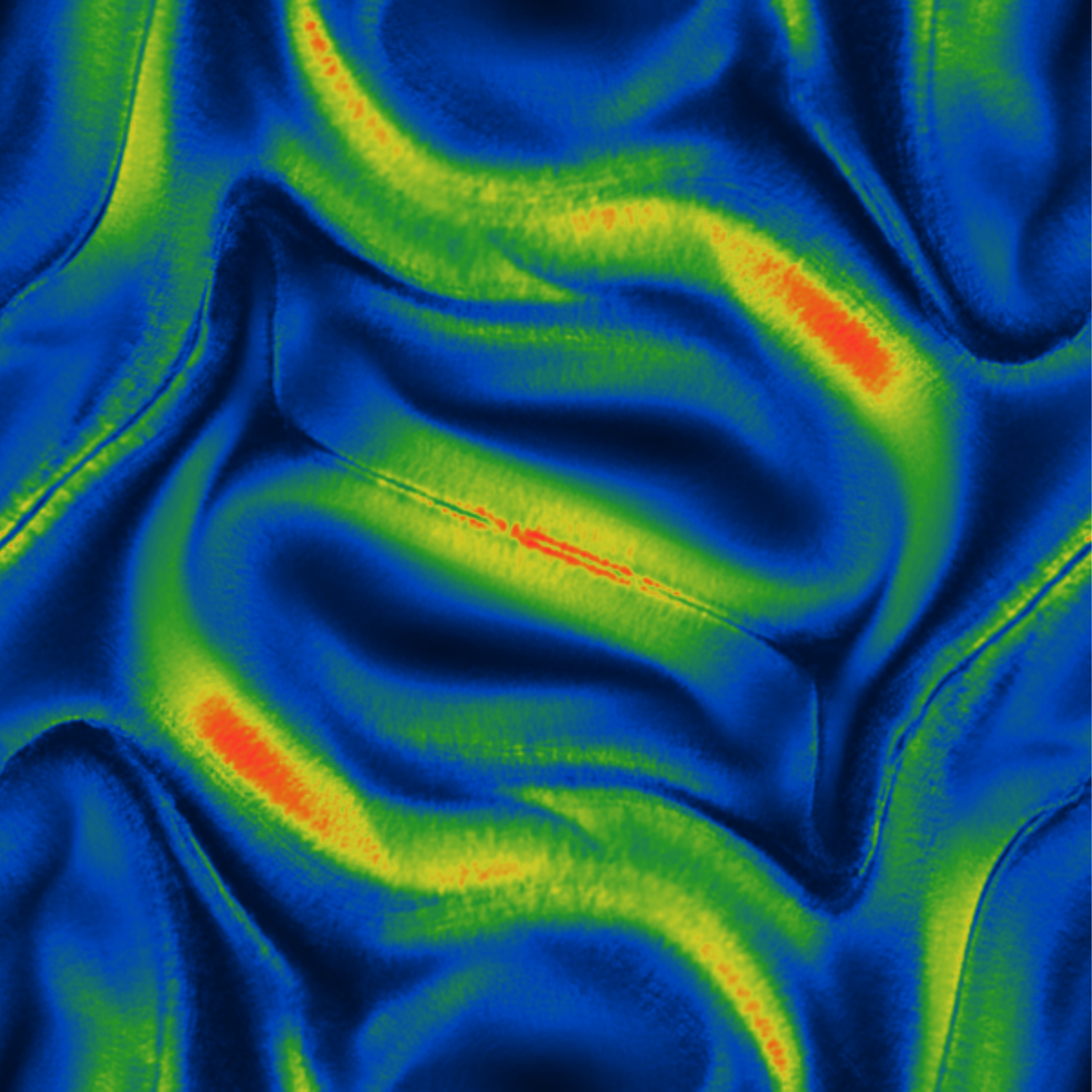}
 & \includegraphics[height=0.22\textwidth]{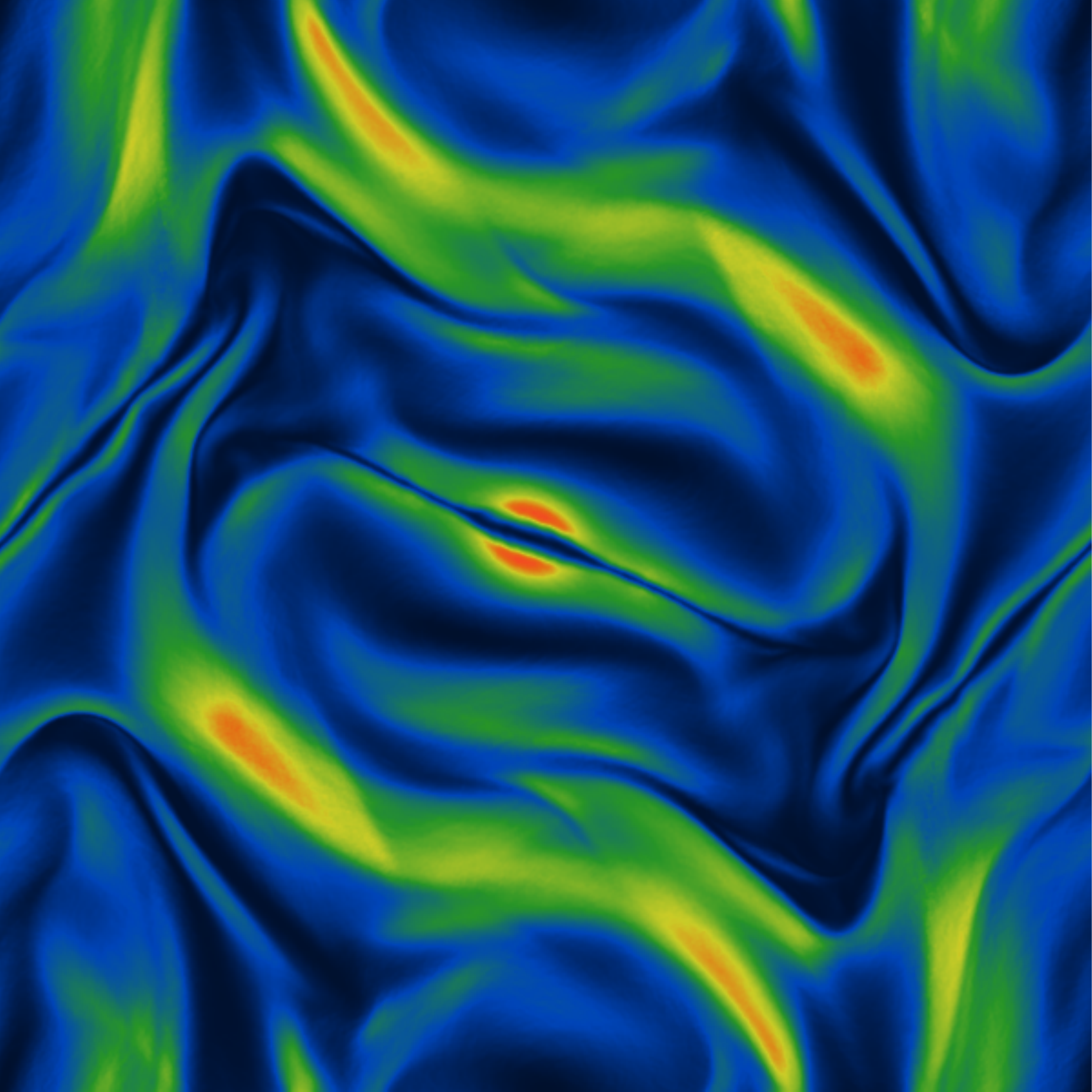}
 & \includegraphics[height=0.22\textwidth]{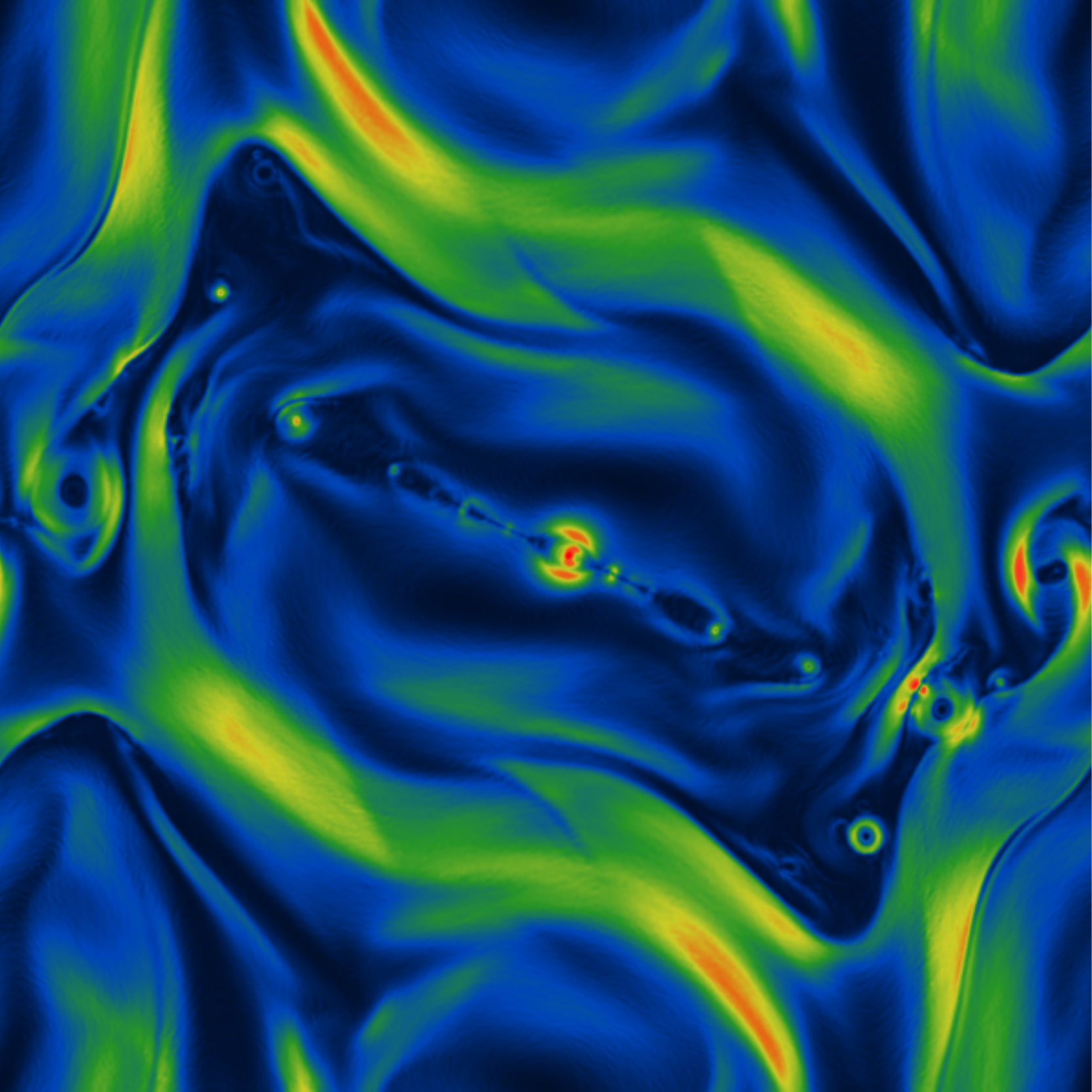}
 & \includegraphics[height=0.22\textwidth]{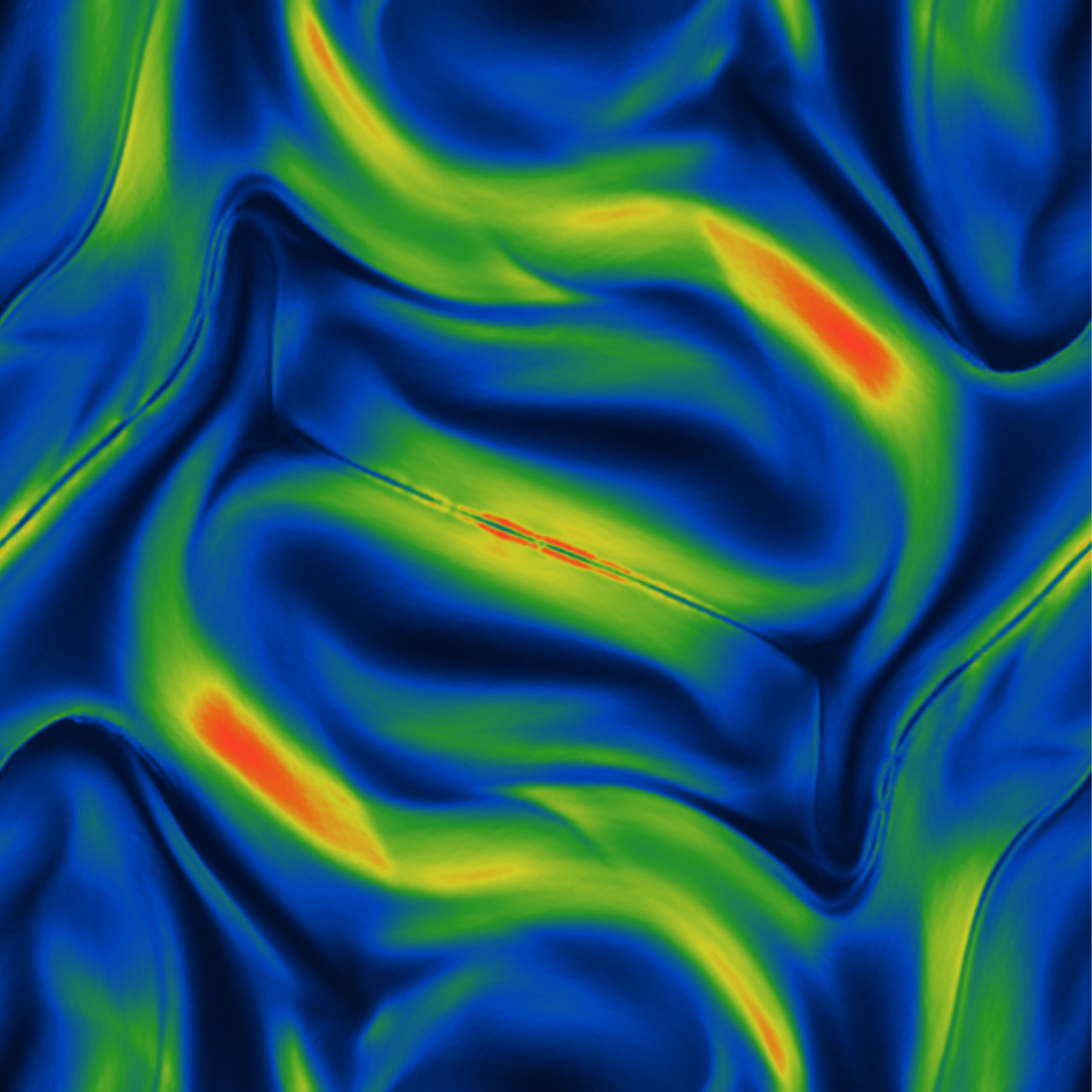}
 & \includegraphics[height=0.22\textwidth]{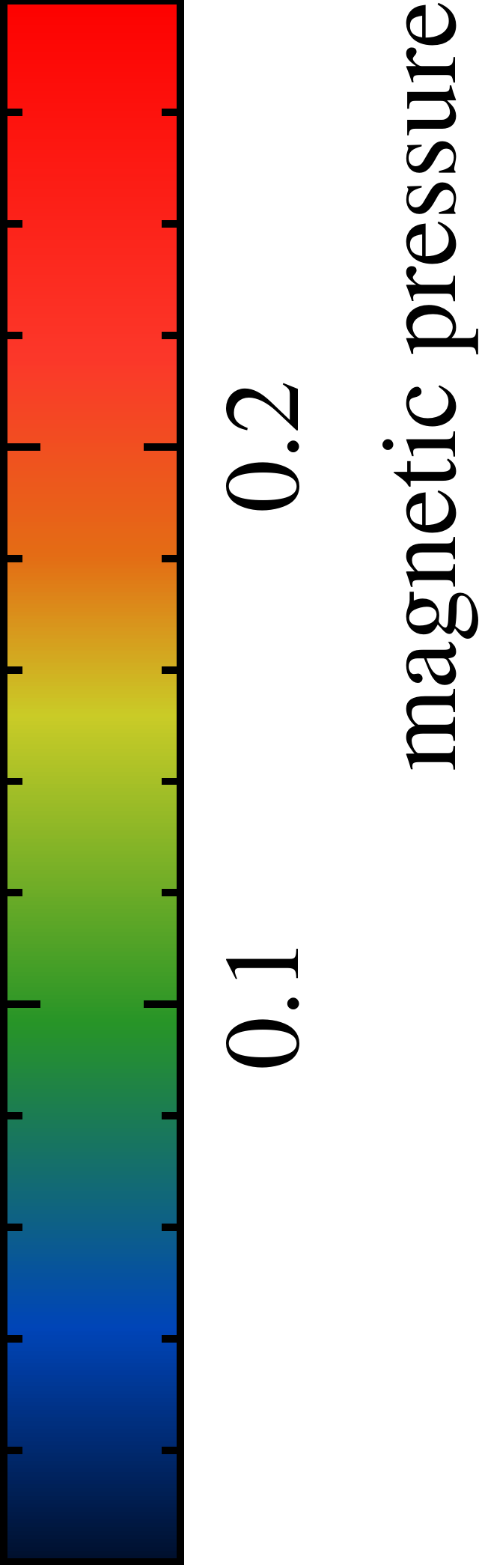}
\\
   \includegraphics[height=0.22\textwidth]{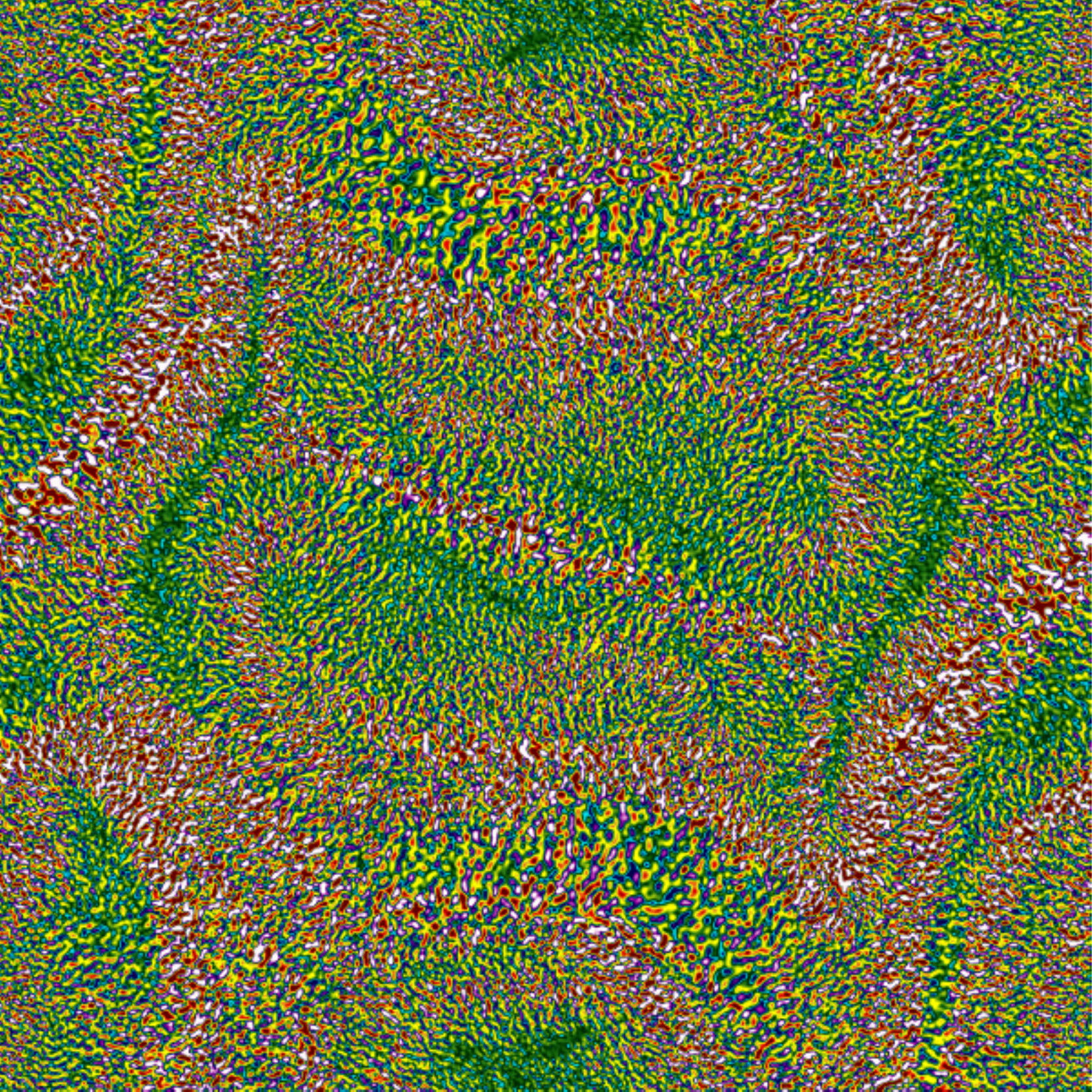}
 & \includegraphics[height=0.22\textwidth]{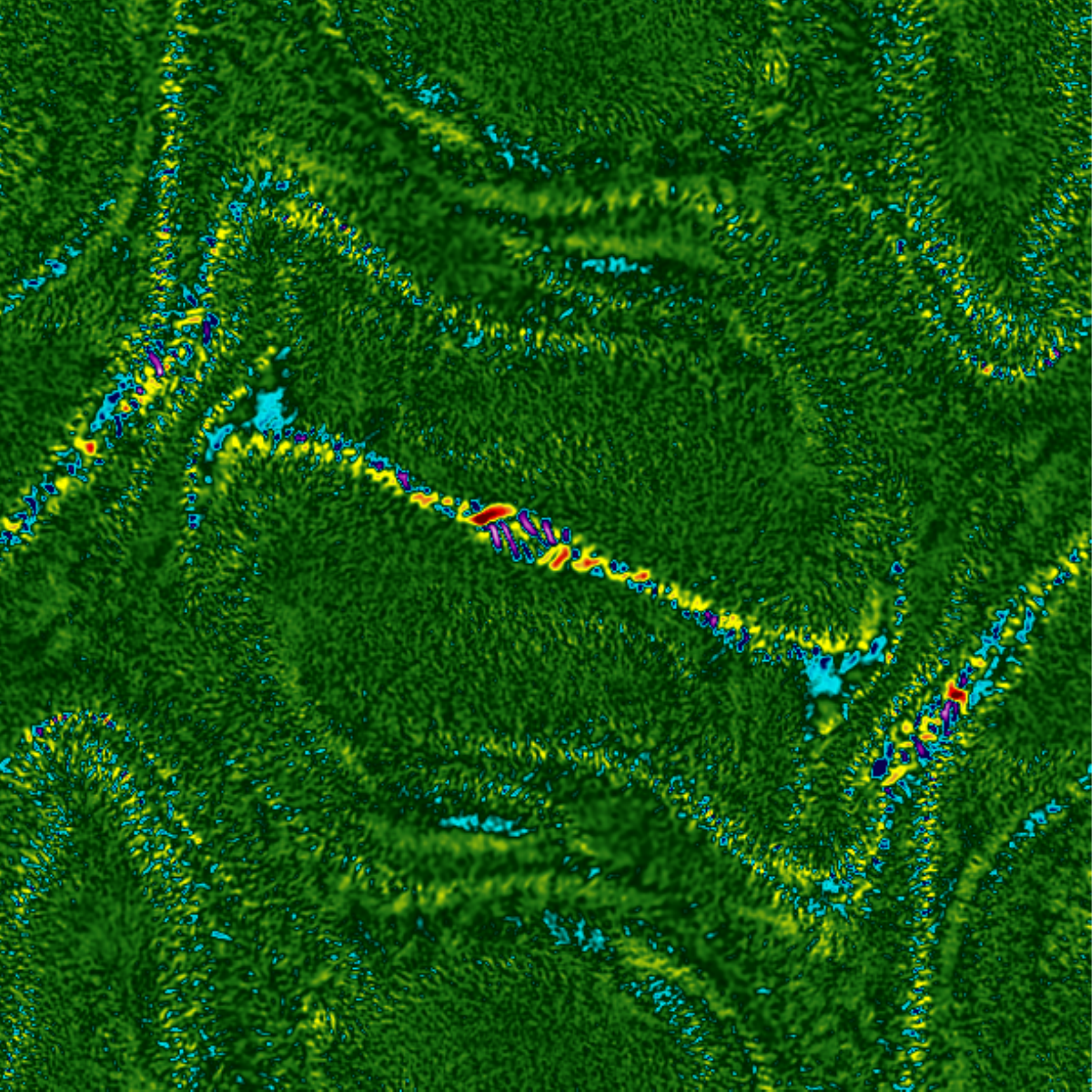}
 & \includegraphics[height=0.22\textwidth]{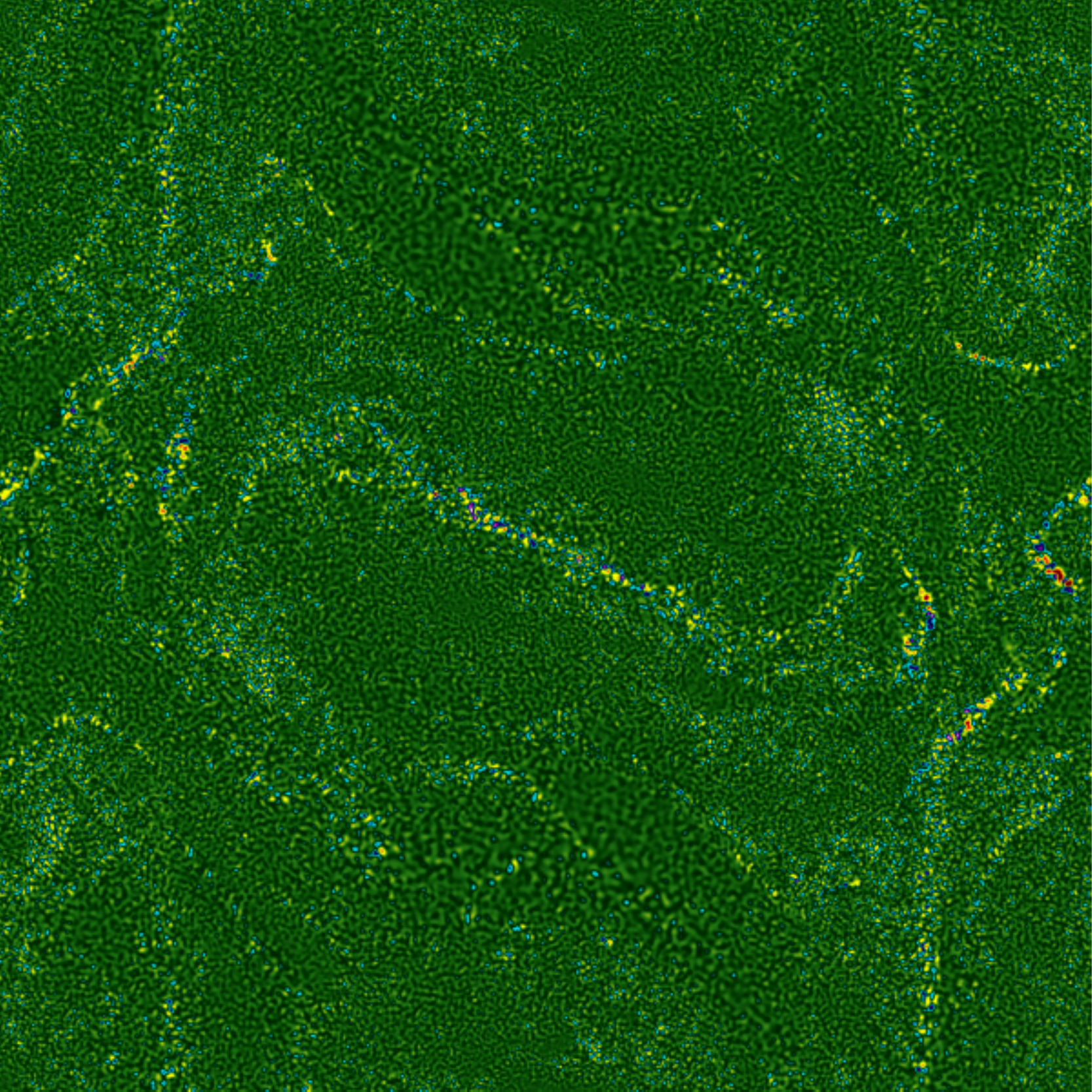}
 & \includegraphics[height=0.22\textwidth]{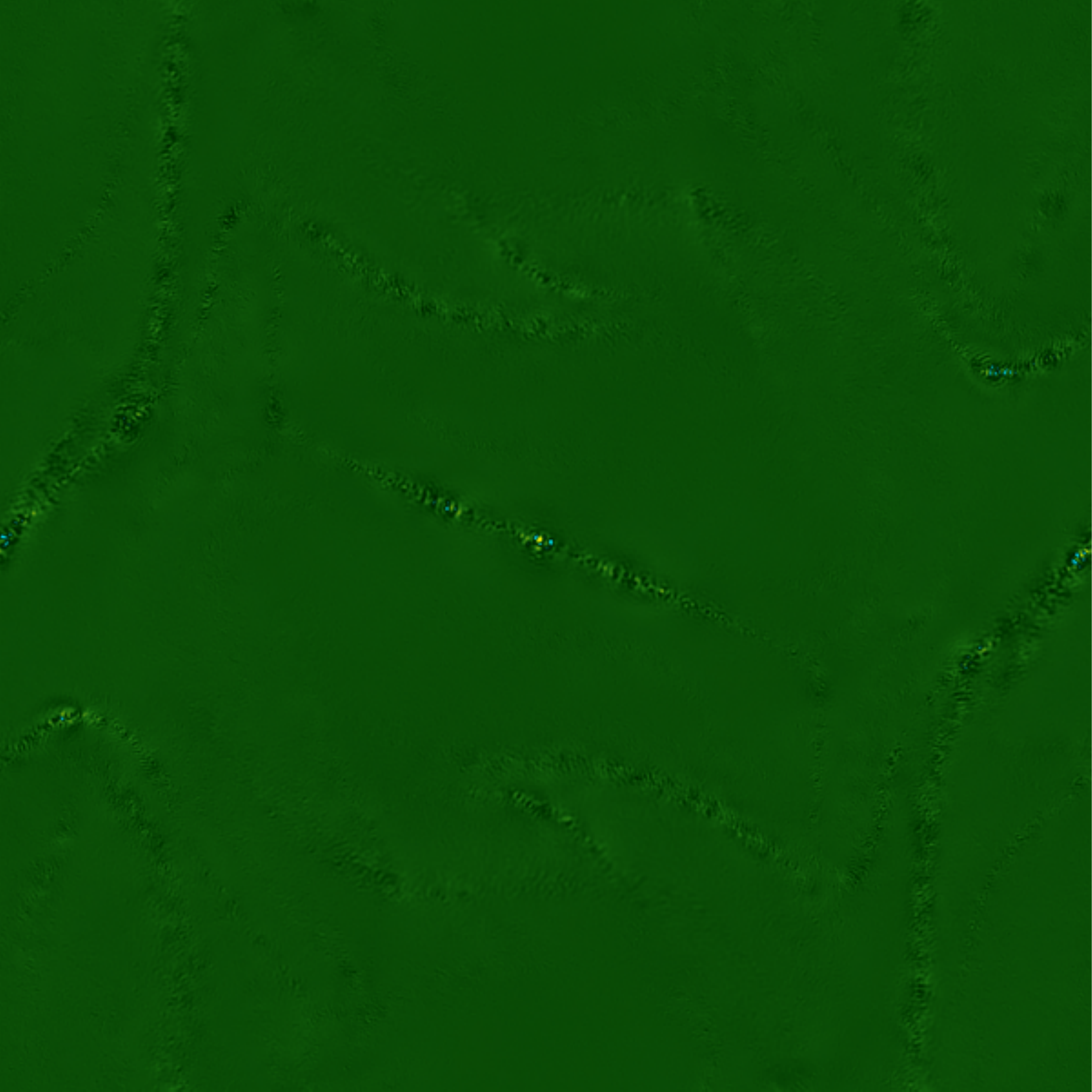}
 & \includegraphics[height=0.22\textwidth]{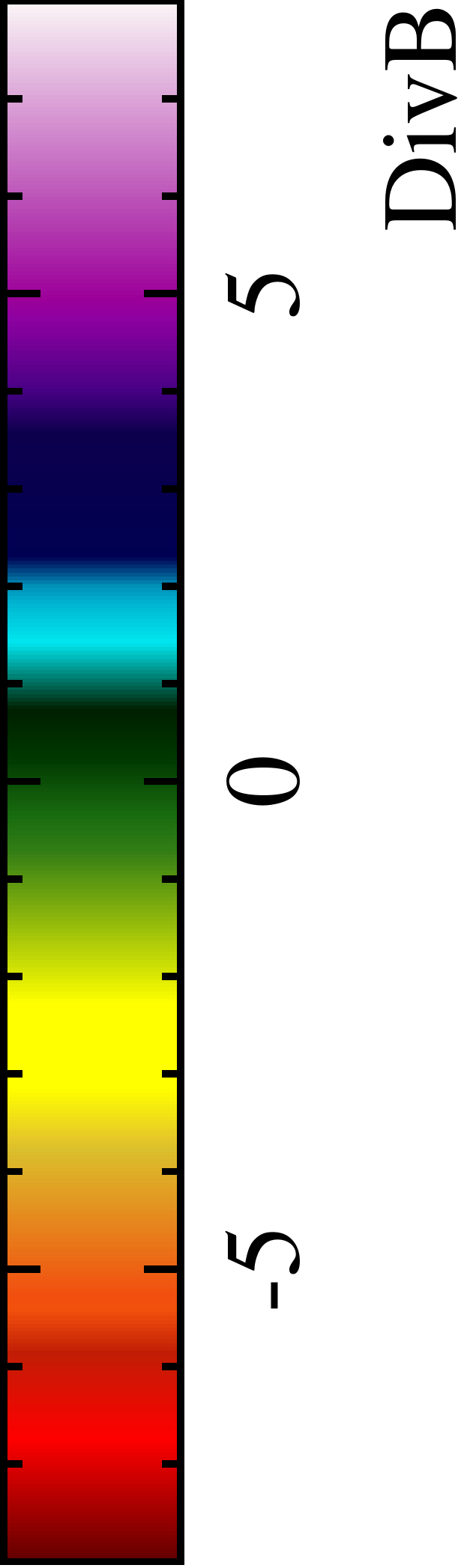}
\end{tabular}
\caption{The density (top row), magnetic pressure (middle row), and divergence of ${\bf B}$ (bottom row) in the Orszag-Tang vortex at $t=1.0$ comparing the control case (far left), including artificial resistivity (centre left), evolving the magnetic field using Euler Potentials (centre right), and applying the constrained divergence cleaning method (far right).}
\label{fig:orszag-compilation}
\end{figure*}

The constrained cleaning methods ability to handle free surfaces is investigated by considering a disc of fluid with open boundary conditions.  For this test, the full SPMHD equations are not solved so that the fluid retains its shape.  Instead, only the cleaning subsystem of equations are utilised.

The fluid is contained within a disc of radius $R=1$ composed of 1976 particles placed on a cubic lattice.  The initial magnetic field is $B_z = 1 / \sqrt{4\pi}$, with a perturbation in the x-component of the field of the form
\begin{equation}
\label{eq:adv-divergence-perturbation}
B_x = \frac{1}{\sqrt{4 \pi}} \left[ \left(r / r_0\right)^8 - 2 \left( r/r_0 \right)^4 + 1 \right]; \hspace{5mm} r < r_{0},
\end{equation}
centred on a region in the middle of the disc of radius \mbox{$r_0 = 1/\sqrt{8}$}.  The density is uniformly $\rho=1$ with zero velocity field.

The maximum divergence error over time for the previous implementation along with the new constrained implementation of divergence cleaning is shown in Fig.~\ref{fig:leftright-div-plots}.  Both cases show undamped (purely hyperbolic) and damped (hyperbolic/parabolic) cleaning.  For the previous implementation, once the divergence waves reach the hard edge of the disc, it causes divergence (and magnetic energy) to increase exponentially.  This behaviour also occurs across jumps in density.  However, the constrained cleaning method models the boundary interaction correctly.

\subsection{Orszag-Tang Vortex}
\label{sec:ot}

The constrained cleaning method is compared against artificial resistivity and Euler Potentials using the Orszag-Tang vortex test problem.  This problem has been widely used as a test of MHD codes because of its complex dynamics, consisting of several classes of interacting shockwaves.  

The problem is set up in a box with dimensions $x,y \in [0,1]$ with periodic boundary conditions.  The initial gas state is set to $\rho = 25 / (36\pi)$, $P = 5/(12\pi)$, $\gamma = 5/3$, with velocity field ${\bf v} = [-\sin(2\pi y), \sin(2\pi x)]$.  The initial magnetic field is ${\bf B} = [-\sin(2\pi y), \sin(4\pi x)]$.  All examples presented use $512 \times 590$ particles initially arranged on a hexagonal lattice.

\begin{figure}
 \centering
\includegraphics[width=\linewidth]{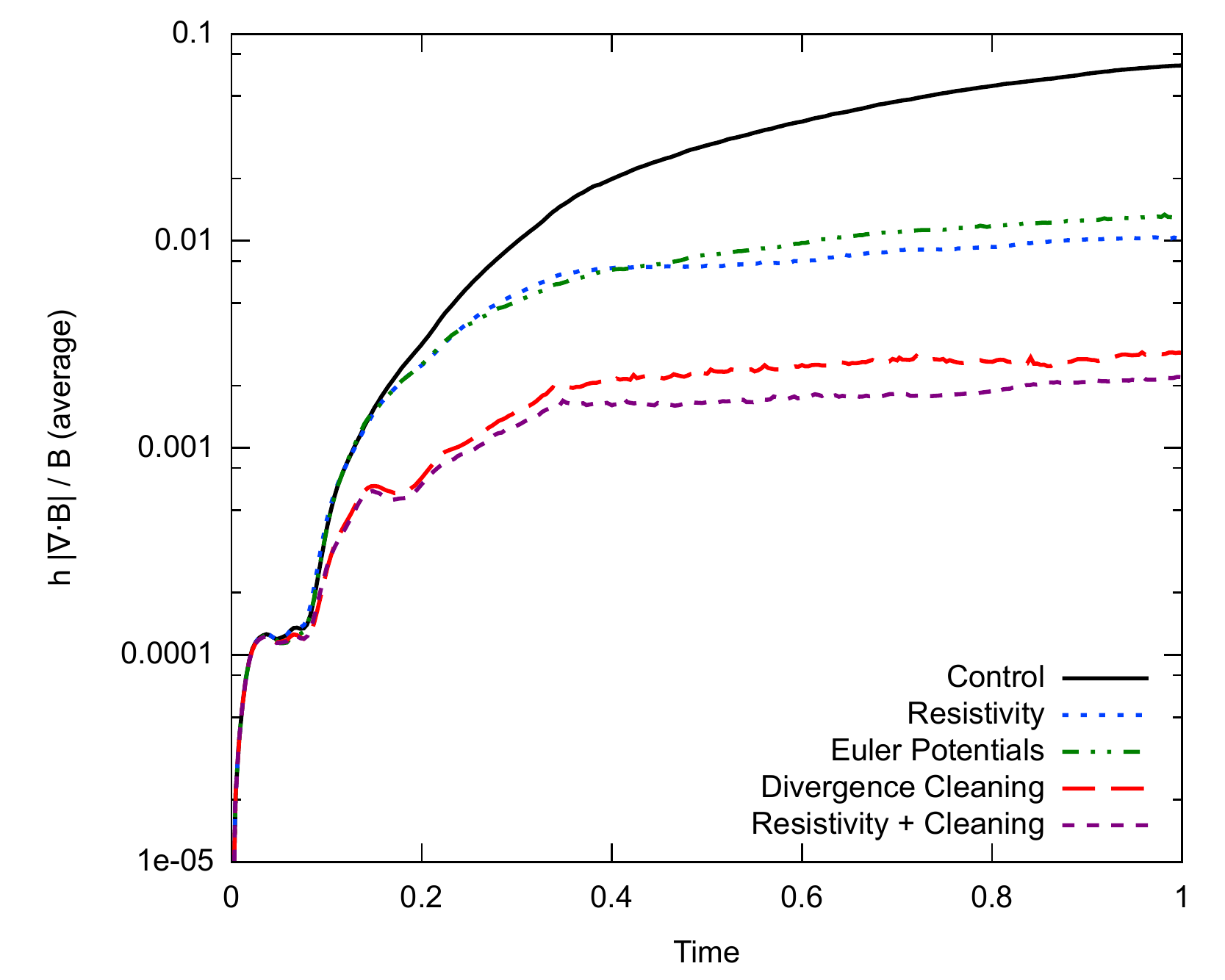}
\caption{Average divergence error as a function of time in the Orszag-Tang vortex.  Test cases included are: no divergence control, using artificial resistivity, employing Euler Potentials, applying divergence cleaning, and divergence cleaning plus resistivity.  Divergence cleaning provides an order of magnitude reduction in divergence error over resistivity or Euler Potentials.}
\label{fig:orszag-divb}
\end{figure}

Results are obtained for five cases: i) no divergence control, ii) artificial resistivity, iii) Euler Potentials, iv) divergence cleaning, and v) divergence cleaning plus resistivity.  Fig.~\ref{fig:orszag-compilation} shows the density, magnetic pressure, and divergence throughout the system at $t=1$ for the first four cases.  Significant divergence is present in the magnetic field for the control case, which is reflected by small disturbances in the density and magnetic pressure.  Artificial resistivity and Euler Potentials have an order of magnitude lower divergence error by comparison (Fig.~\ref{fig:orszag-divb}).  Applying divergence cleaning produces significantly improved results.  Divergence throughout the system is negligible, with average divergence error reduced by two orders of magnitude in comparison to the control case, down to $\sim0.1\%$.


\subsection{Gravitational collapse of a magnetised molecular cloud core}
\label{sec:star}

\begin{figure}
\centering
\includegraphics[width=\linewidth]{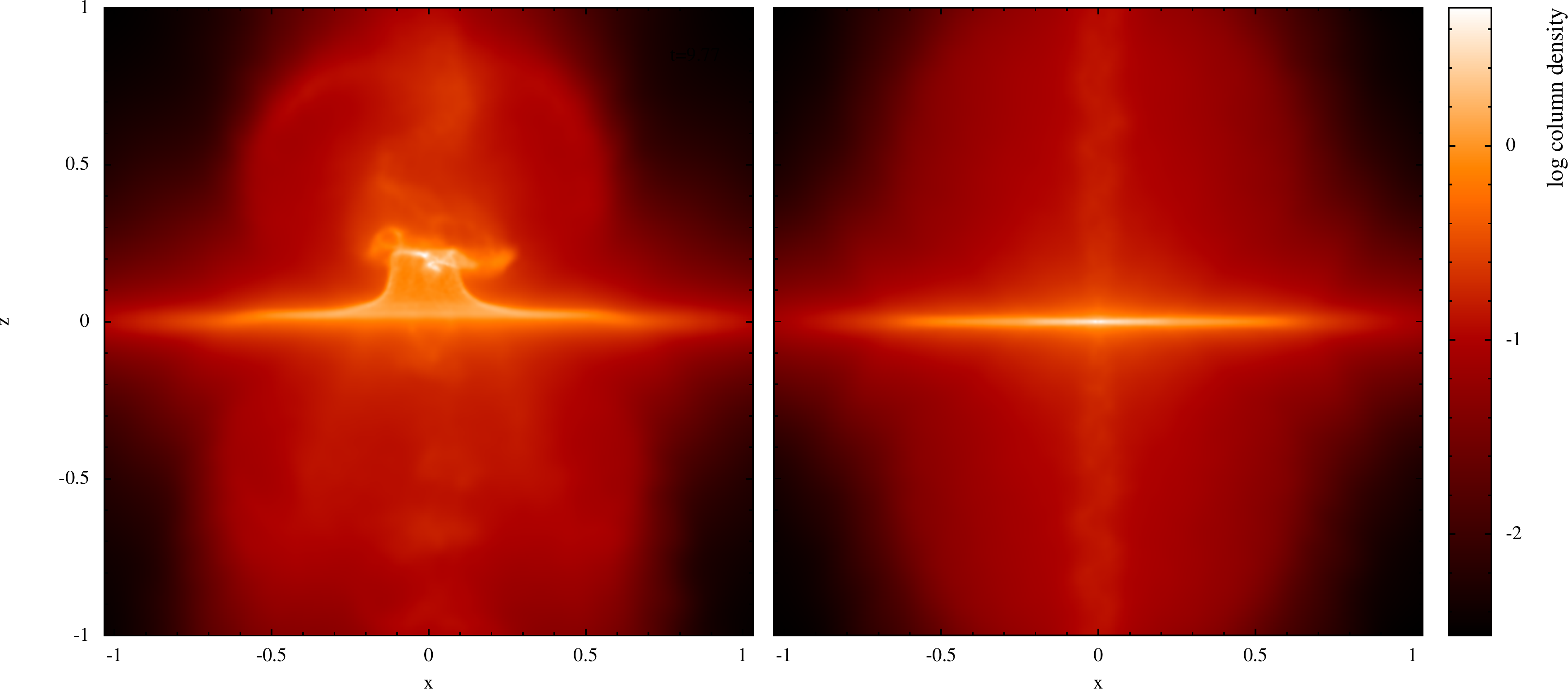}
\caption{Column density along the y-axis of the star formation problem at $t=1.1 t_\text{ff}$.  The majority of the gas has been flattened to form an accretion disk about the protostar.  In the left panel, magnetic divergence has grown too large and disrupted the system.  In the right panel, divergence cleaning has been applied, stabilising the evolution of the system.}
\label{fig:star-column-density}
\end{figure}

\begin{figure}
\centering
\includegraphics[width=\linewidth]{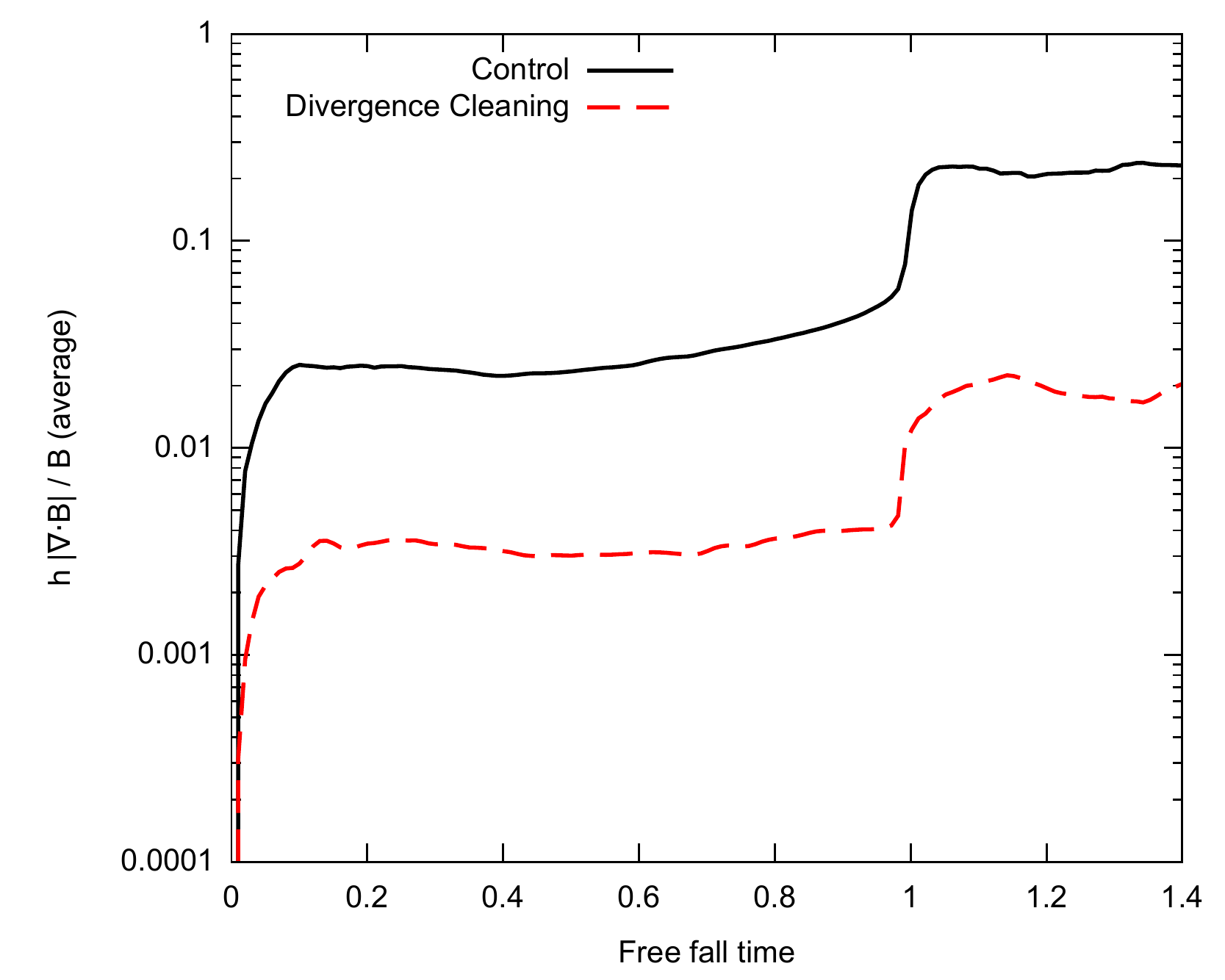}
\caption{Average divergence error as a function of time for the star formation problem.  Applying divergence cleaning reduces error by an order of magnitude.}
\label{fig:star-divb}
\end{figure}

\begin{figure}
\includegraphics[width=\linewidth]{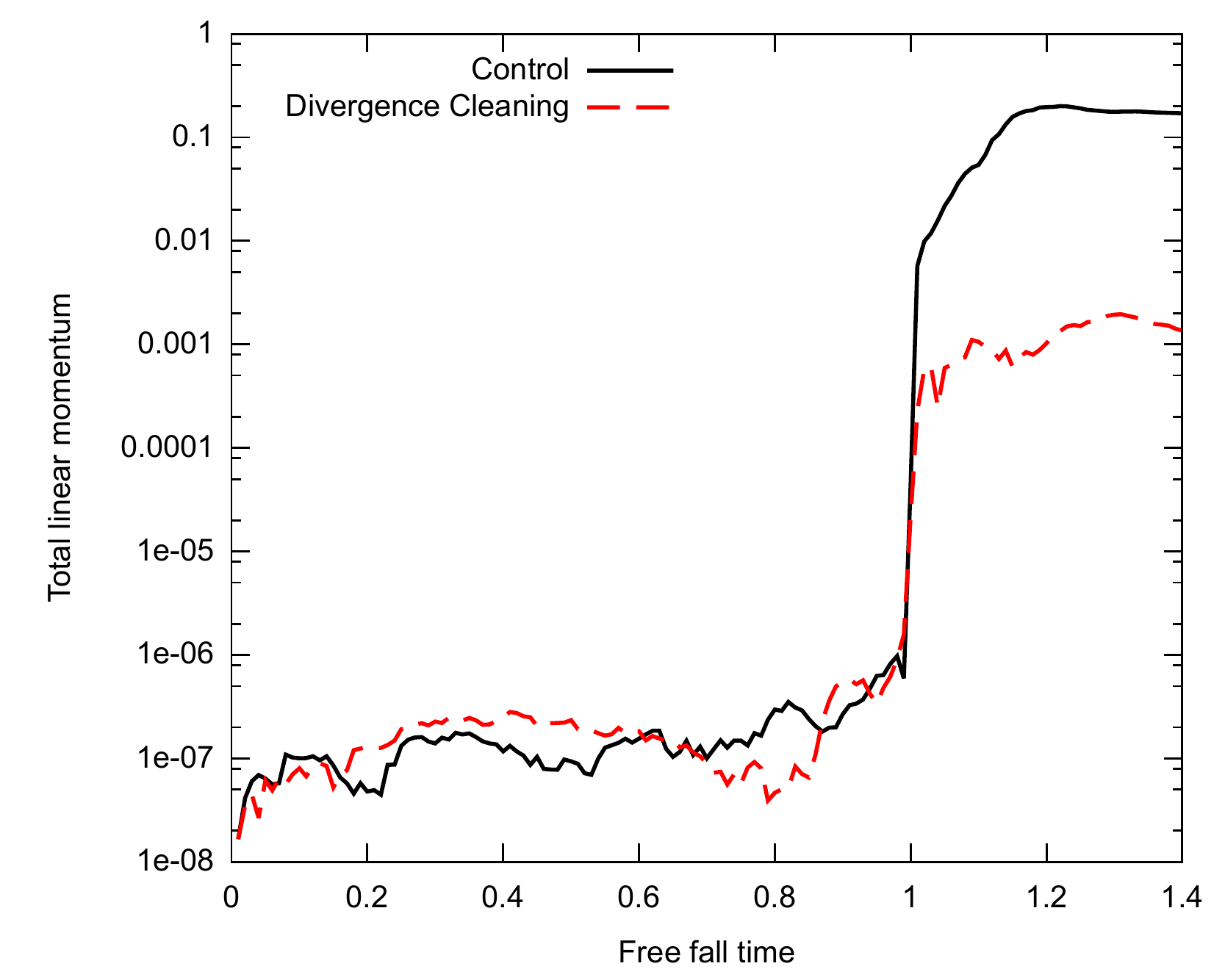}
\caption{Total linear momentum for the star formation problem.  Once the collapse reaches peak density ($t\sim1$), a sharp increase in momentum occurs due to divergence errors.  The divergence cleaned system reduces this momentum spike by two orders of magnitude.}
\label{fig:star-mom}
\end{figure}

Our final test is drawn from our intended application: simulations of star formation that involve magnetic fields \cite{jetletter}.  These simulations follow \cite{2007MNRAS.377...77P}, where an initial one solar mass sphere of gas with uniform magnetic field in the $z$-direction and in solid body rotation contracts under self-gravity to form a protostar with surrounding disc.  However, at times near peak density, the magnetic field in the dense central region becomes strong and can produce high divergence errors.  This has limited the range of initial magnetic field strengths which could be simulated, as if the divergence grows too large, the tensile instability correction term (\ref{eq:tensile-instability-correction}) injects enough momentum into the system to erroneously eject the protostar out of its disc \cite{pf10b}.  Thus, this simulation proves an excellent demonstration of the capabilities of the constrained hyperbolic divergence cleaning method to reduce divergence errors in realistic, 3D simulations.

\begin{figure*}
 \centering
\setlength{\tabcolsep}{0.005\textwidth}
\setlength\fboxsep{0pt}
\setlength\fboxrule{0.5pt}
\begin{tabular}{ccccl}
   \fbox{\includegraphics[height=0.22\textwidth]{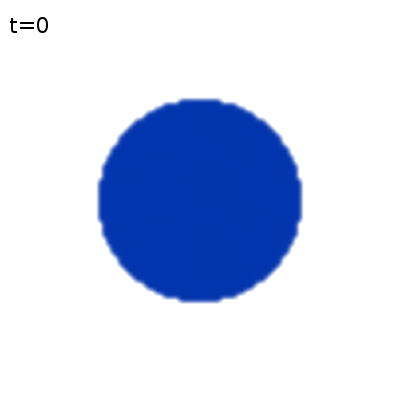}}
 & \fbox{\includegraphics[height=0.22\textwidth]{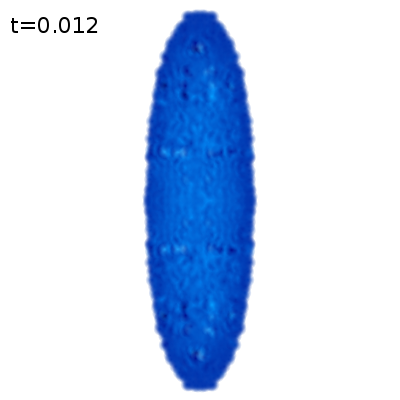}}
 & \fbox{\includegraphics[height=0.22\textwidth]{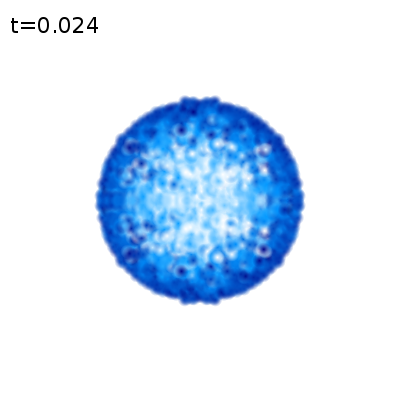}}
 & \fbox{\includegraphics[height=0.22\textwidth]{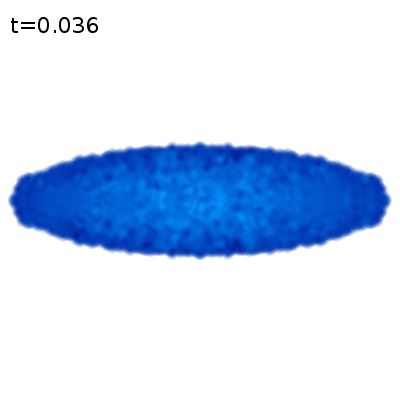}}
 & \includegraphics[height=0.22\textwidth]{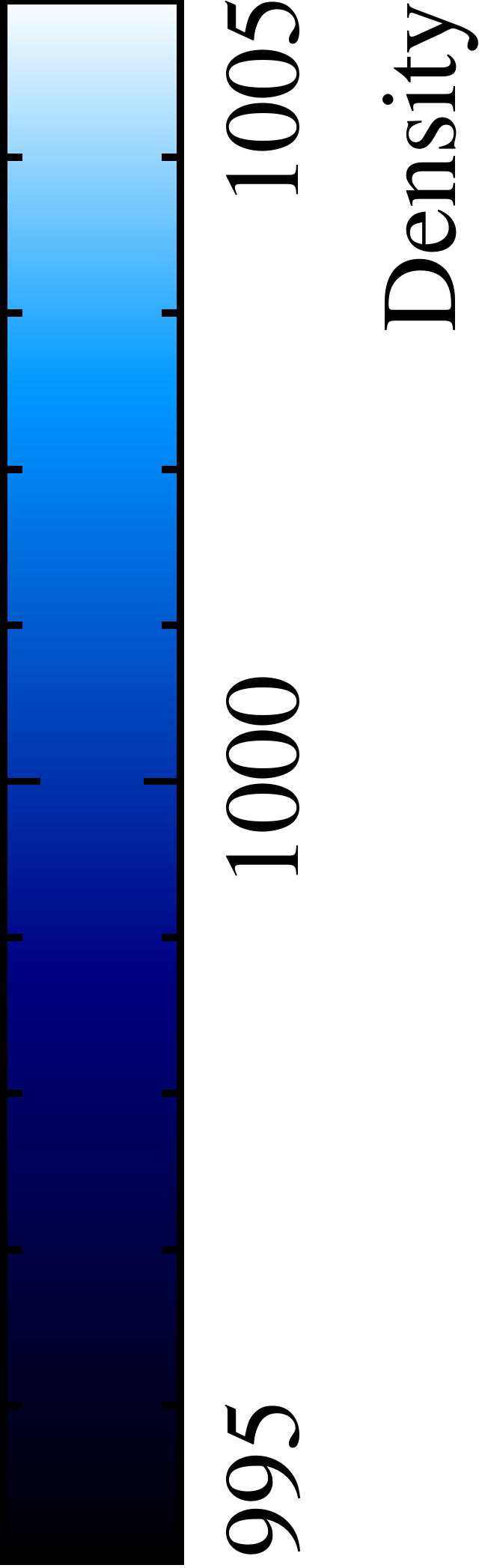}
\end{tabular}
\caption{Snapshots of the oscillating water drop test.  The circular drop has an initial velocity which squeezes it into an elliptical shape along the $y$-axis.  A radial force is present which halts the expansion of the drop, then contracts it to its original shape before expanding along the opposite axis.  This behaviour repeats causing the drop to oscillate alternately along the two axes.}
\label{fig:drop-compilation}
\end{figure*}

The sphere of gas has radius $R = 4\times10^{16} \text{cm}$ with uniform density $\rho = 7.43\times10^{-18}$ g $\text{cm}^{-3}$.  A barotropic equation of state is used, as described in \cite{2007MNRAS.377...77P}.  The magnetic field strength is set to give a mass-to-magnetic flux ratio of 5 times the critical value for magnetic fields to provide support against gravitational collapse.  To avoid edge effects with the magnetic field, the sphere is embedded in a periodic box of length $4R$ containing material surrounding the sphere set in pressure equilibrium with density ratio 1:30.  This test uses only a minimal amount of resistivity, with $\alpha_B \in [0, 0.1]$. Self-gravity is simulated using a hierarchical partitioning tree, with gravitational force softening using the SPH kernel as described by \cite{pm07}.  The free fall time is $\sim 24000$ years.  A sink particle is inserted once the gas density surpasses $\rho_{\text{sink}} = 10^{-10}$ g $\text{cm}^{-3}$, and accretes particles within a radius of $6.7$ AU.

Fig.~\ref{fig:star-column-density} shows column density comparisons of simulations with (right) and without (left) divergence cleaning at $t=1.1$ free fall time, showing that drastic improvements to the results are obtained by incorporating divergence cleaning.  The protostar remains stable in its disc and a helical shaped jet is launched from the centre \cite{jetletter}.  The average divergence error is reduced by an order of magnitude (Fig.~\ref{fig:star-divb}), and this leads to a corresponding improvement in the momentum conservation of roughly two orders of magnitude (Fig.~\ref{fig:star-mom}).

\subsection{Oscillating water drop test}
\label{sec:drop}

To investigate the effectiveness of our velocity cleaning algorithm, it is applied to an oscillating elliptic water drop.  The water drop is initially circular and is free standing.  A radial force is exerted upon it, and with an initial velocity which is compressional along one axis, the drop oscillates, squeezing alternately along the two axes.  This behaviour is demonstrated in Fig.~\ref{fig:drop-compilation}.

The drop is modelled using the weakly compressible approximation (equations (\ref{eq:wcsph-momentum}) and (\ref{eq:wcsph-continuity}) with (\ref{eq:wcsph-eos}) as the equation of state).  The reference density is $\rho_0 = 1000$ kg $\text{m}^{-2}$, and the initial velocity field is ${\bf v} = [-100x, 100y]$.  The radial force is $-100^2 {\bf r}$.  The drop has radius $R=1$, and a total of $1976$ particles are used arranged on a square lattice.

\begin{figure}
 \centering
\includegraphics[width=\linewidth]{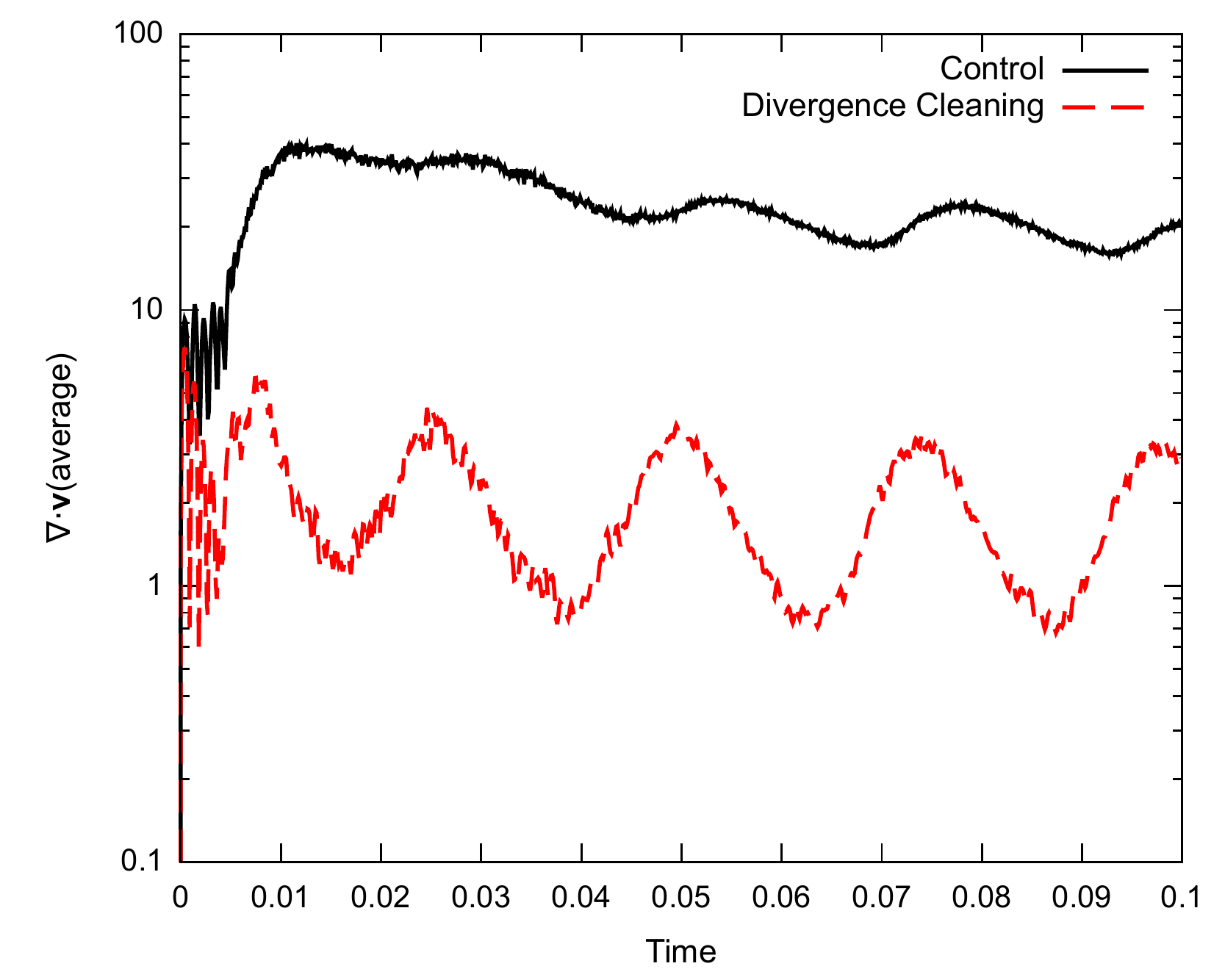} 
\caption{Average $\nabla \cdot {\bf v}$ of the elliptic water drop test.  Average velocity divergence is reduced by approximately an order of magnitude when divergence cleaning is applied.}
\label{fig:drop-avgdiv}
\end{figure}

\begin{figure}
\includegraphics[width=\linewidth]{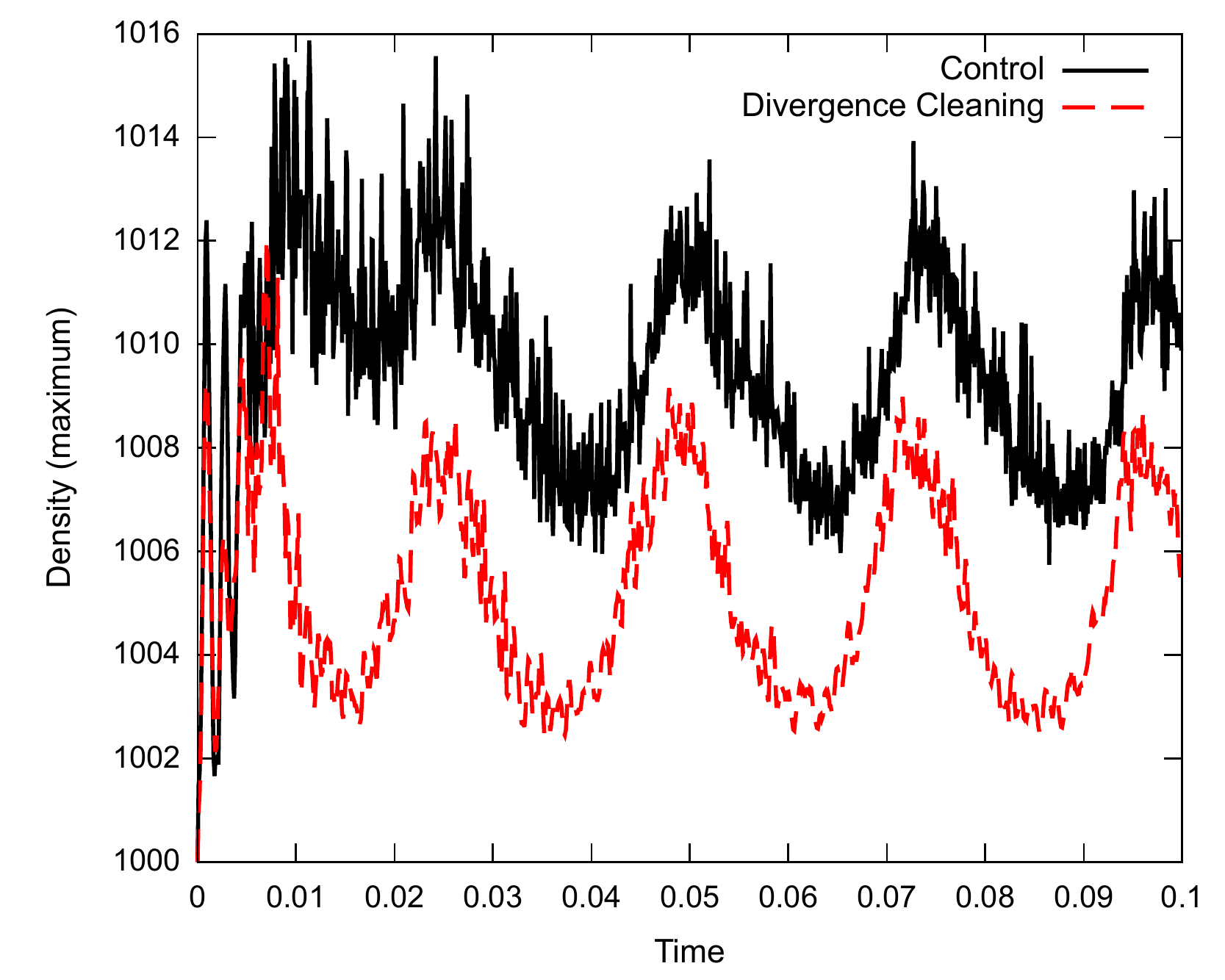}
\caption{Maximum density variation during the elliptic water drop test.  Applying divergence cleaning to the velocity field reduces density changes from the reference density by $\sim 0.5$.}
\label{fig:drop-maxrho}
\end{figure}

\begin{figure}
 \includegraphics[width=\linewidth]{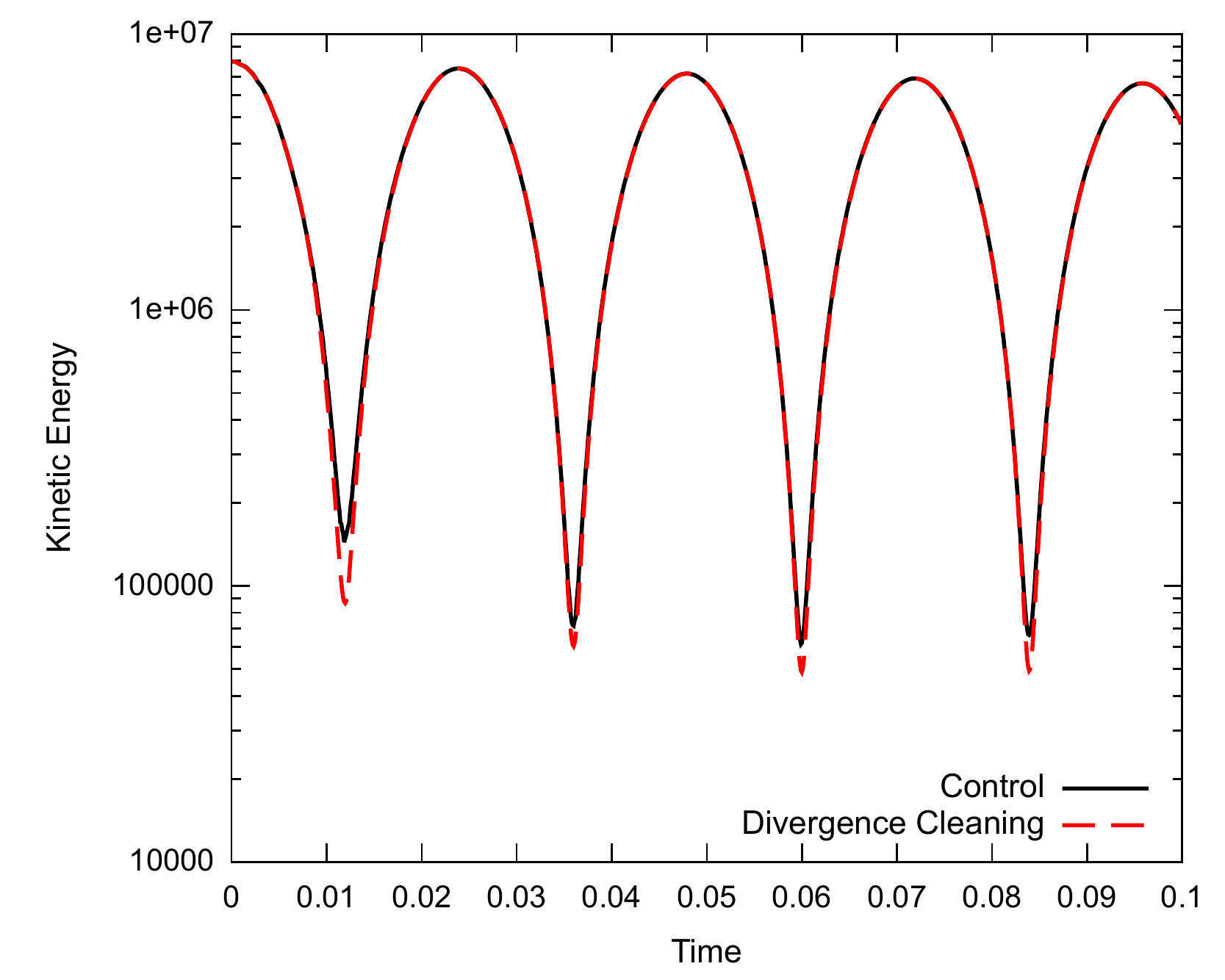}
\caption{Total kinetic energy of the elliptic water drop test.  No significant discrepancies exist between the control and divergence cleaned tests.}
\label{fig:drop-ke}
\end{figure}

The evolution of the drop is tracked until $t=0.1$, approximately two oscillation periods.  Fig.~\ref{fig:drop-avgdiv} shows the average velocity divergence of the system as a function of time for both the cleaned and uncleaned systems.  Applying cleaning reduces the average divergence by nearly an order of magnitude, similar to results obtained for magnetic field cleaning.  This leads to a reduction in maximum density error by a factor of two (Fig.~\ref{fig:drop-maxrho}).  The dissipation of kinetic energy by the cleaning algorithm is insignificant, as shown in Fig.~\ref{fig:drop-ke}.

\section{Conclusion}
\label{sec:conclusion}

In this paper, SPH formulations of Dedner et al's hyperbolic/parabolic divergence cleaning for the magnetic and velocity fields have been presented.  The algorithm is attractive because it is computational inexpensive and easy to implement in existing codes.  For SPMHD simulations in particular, it represents a path forward for maintaining the magnetic divergence constraint without the drawbacks associated with using Euler Potentials or artificial resistivity.  

Our method was derived by considering the energy contained in the $\psi$ field as part of the system total energy.  With this contribution included, it is possible to construct SPH implementations which conserve energy and, in the case of the velocity field, momentum.  This stabilises the algorithm across density jumps and at free surface boundaries. Results obtained find an order of magnitude reduction in average level of divergence for all tests of magnetic and velocity field cleaning.  For 3D astrophysical star formation problems, this reduction in magnetic divergence has improved momentum conservation by two orders of magnitude.  When the velocity field is cleaned in weakly compressible SPH simulations of an oscillating water drop, density errors are reduced by half with negligible kinetic energy dissipation.

Given the performance of the algorithm on complicated astrophysical applications, the door to study other astrophysical problems is now open.  Though the results of velocity cleaned weakly compressible SPH are encouraging, additional work is required, in particular, for cases involving boundary particles.

\section*{Acknowledgment}
The authors thank Matthew Bate and Joe Monaghan for useful discussions. T. Tricco is supported by Endeavour IPRS and APA postgraduate research scholarships. DJP acknowledges support from the Australian Research Council via Discovery Project grant DP1094585. We acknowledge the use of \textsc{splash}/\textsc{giza} \cite{splashpaper}.



%

\bibliographystyle{IEEEtran}
\bibliography{spheric-bib}

%
%

\end{document}